\documentclass[a4paper,fleqn,usenatbib]{mnras}

\usepackage{newtxtext,newtxmath}
\usepackage[T1]{fontenc}
\usepackage{ae,aecompl}
\usepackage{graphicx}	
\usepackage{enumitem}
\usepackage{rotating}
\usepackage{pdflscape}

\newcommand{\sw}{\emph{Swift}}
\newcommand{\ch}{\emph{Chandra}}
\newcommand{\xmm}{\emph{XMM-Newton}}

\newcommand{\gc}{$\gamma$\,Cas}
\newcommand{\kms}{km\,s$^{-1}$}

\title[X-rays from Be+sdO]{The X-ray emission of Be+stripped star binaries\thanks{Based on spectra obtained with \xmm, \sw, and \ch }}

\author[Y. Naz\'e et al.]{Ya\"el~Naz\'e$^1$\thanks{F.R.S.-FNRS Senior Research Associate, email: ynaze@uliege.be}, Gregor Rauw$^1$, Myron A. Smith$^2$, Christian Motch$^3$ 
\\
$^1$ Groupe d'Astrophysique des Hautes Energies, STAR, Universit\'e de Li\`ege, Quartier Agora (B5c, Institut d'Astrophysique et de G\'eophysique), \\
All\'ee du 6 Ao\^ut 19c, B-4000 Sart Tilman, Li\`ege, Belgium\\
$^2$ NSF OIR Lab, 950 N Cherry Ave, Tucson, AZ 85721, USA\\
$^3$ Universit\'e de Strasbourg, CNRS, Observatoire Astronomique de Strasbourg, 11 rue de l'Universit\'e, UMR 7550, F-67000 Strasbourg, France
}

\pubyear{2021}

\begin{document}
\label{firstpage}
\pagerange{\pageref{firstpage}--\pageref{lastpage}}
\maketitle

\begin{abstract}
Using observations from \ch, \sw\ and \xmm, we investigate the high-energy properties of all known (18) Be+sdO systems as well as 7 additional Be binaries suspected to harbour stripped stars. The observed X-ray properties are found to be similar to those observed for other Be samples. The vast majority of these systems (15 out of 25) display very faint (and soft) X-ray emission, and six others are certainly not bright X-ray sources. Only two systems display \gc\ characteristics (i.e. bright and hard X-rays), and one of them is a new detection: HD\,37202 ($\zeta$\,Tau). It presents an extremely hard spectrum, due to a combination of high temperature and high absorption (possibly due to its high inclination). In parallel, it may be noted that the previously reported cyclic behaviour of this Be star has disappeared in recent years. Instead, shorter cycles and symmetric line profiles are observed for the H$\alpha$ line. It had been recently suggested that the peculiar X-ray emissions observed in \gc\ stars could arise from a collision between the disk of a Be star and the wind of its hot, stripped-star companion. The small fraction of \gc\ analogs in this sample, as well as the properties of the known companions of the \gc\ cases (low mass or not extremely hot, contrary to predictions), combined to the actual stripped-star and colliding-wind empirical knowledge, make the disk-wind collision an unlikely scenario to explain the \gc\ phenomenon.
\end{abstract}


\section{Introduction}
While nearly all types of stars have been detected to emit X-rays, the high-energy emissions from most massive stars could be linked up to now to a single cause: their stellar winds. Indeed, the line-driven winds are intrinsically unstable, so shocks naturally arise between different wind parcels, leading to X-ray emission \citep{fel97}. Such X-rays have been recorded for tens of OB stars and their properties are well known \citep[e.g.][]{ber96,osk05,san06,naz11,rau15b}: they are soft ($kT\sim0.6$\,keV on average) and rather faint ($\log(L_{\rm X}/L_{\rm BOL})\sim -7$). In addition, when a strong dipolar magnetic field is present, as occurs in $\sim$7\% of OB stars \citep{fos15,wad16}, it channels the wind flows from both stellar hemispheres towards the magnetic equator where they collide, generating additional X-rays \citep[for a review, see][]{udd16}. Also, the winds of the two components in a massive binary system can collide, which may generate X-rays in some systems \citep[for a review, see][]{rau16}. In both cases, this additional X-ray production leads to an enhanced luminosity (up to $\log(L_{\rm X}/L_{\rm BOL})\sim -6$) with an emission often harder in nature. 

Up to now, this general portrait did not include the peculiar \gc\ category. Stars classified as \gc\ analogs all belong to the Oe/Be category, i.e. they display Balmer emission lines associated with the presence of a Keplerian circumstellar decretion disk. Furthermore, they all emit peculiar X-rays, at odds with the properties recorded for other OB stars. Indeed, the X-ray emission comes from very hot ($kT>5$\,keV) plasma, is `flaring' on very short timescales, and appears very bright ($\log(L_{\rm X})\sim31.6-33.2$, $\log(L_{\rm X}/L_{\rm BOL})$ of --6.2 to --4) although less extreme in character than for Be X-ray binaries \citep{smi16,naz18,naz20}. 

Recently, \citet{lan20} proposed that this \gc\ emission too was linked to stellar winds, although not those of the Be stars. These authors advocated for an X-ray generation in a collision between the Be disk and the wind of a stripped-star companion. Such a pair arises following binary interactions in many current models aimed at explaining the origin of Be stars. In this context, the companion was initially the most massive star of the system but, as it evolved, it transferred mass to the other star. The gain in mass and angular momentum for the latter object following the interaction transforms it into a Be star while the companion, stripped of its envelope (which explains the nickname ``stripped-star''), is burning helium in its core (explaining the alternative name of ``helium star''). The hot surface of the companion, combined to its low luminosity, makes it appear as a O-type subdwarf (sdO) or, for the rare cases with very high mass-loss rates, with a quasi Wolf-Rayet (qWR) spectrum \citep{got18}. 

\begin{landscape}
  \begin{table}
  \scriptsize
  \caption{List of the Be+sdO systems and candidates, along with their general properties as well as those of the Be star and its companion, if known.}
  \label{opt}
  \begin{center}
  \begin{tabular}{llccccccccccccc}
    \hline
    HD	&Alt. name& sp. type(Be) & $d$ & $E(B-V)$ & $T_{eff}$(Be) & $R$(Be)	  & $M$(Be)    & $\log(L_{\rm BOL}$ & $P_{orb}$ & $i$ & $T_{eff}$(comp) & $R$(comp)   & $M$(comp) & $\log(L_{\rm BOL}$ \\
        &         &          &          (pc)& & (kK)         & (R$_\odot$) & (M$_\odot$) &           $/L_\odot)$(Be) &  (d)     & ($^{\circ}$) & (kK)          & (R$_\odot$) &(M$_\odot$)& $/L_\odot)$(comp) \\
\hline
\multicolumn{15}{l}{\it Confirmed Be+sdO sample}\\
HD\,10516 & $\phi$\,Per   & B1.5Ve$^{m15}$  & 184$\pm$8  & 0.162$^{t13}$ & 29.3$^{m15}$ & 4.6$\pm$0.4$^{m15}$  & 9.6$\pm$0.3$^{m15}$ & 4.16$\pm$0.04 & 126.6982$^{m15}$        & 77.6$^{m15}$ & 53$^{m15}$  & 0.93$\pm$0.09$^{m15}$ & 1.2$\pm$0.2$^{m15}$& 3.79$\pm$0.13$^{m15}$  \\
HD\,29441 & V1150\,Tau    & B2.5Vne$^{w21}$ & 622$\pm$17 & 0.198$^{w21}$ & 20.35$^{w21}$& 4.51$\pm$0.32$^{w21}$&                    & 3.47$\pm$0.02 &                        &             & 40$^{w21}$  &  0.24$\pm$0.03$^{w21}$&                   & 2.11$_{-0.21}^{+0.14,w21}$\\
HD\,41335 & HR2142        &B1.5IV-Ve$^{p16}$& 507$\pm$61 & 0.1$^s$      & 21$^{p16}$   & 5$^{p16}$            &$\sim10^{p16}$       & 4.17$\pm$0.10 &  80.913$^{p16}$         & 85$^{p16}$   &$>43^{p16}$  & $>$0.13$^{p16}$        &$\sim$0.7$^{p16}$  & $>$1.7$^{p16}$         \\
HD\,43544 & HR2249        & B3V$^{w21}$     & 307$\pm$6  & 0.155$^{w21}$ & 21.5$^{w21}$ & 4.34$\pm$0.38$^{w21}$& 8.5$^{w21}$         & 3.55$\pm$0.02 &                        &             & 38.2$^{w21}$& 0.50$\pm$0.08$^{w21}$ &                   & 2.68$_{-0.23}^{+0.15,w21}$ \\
HD\,51354 & QY\,Gem       & B3Ve$^{w21}$    & 540$\pm$17 & 0.172$^{w21}$ & 20$^{w21}$   & 4.52$\pm$0.29$^{w21}$&                    & 3.49$\pm$0.03 &                        &             & 43.5$^{w21}$& 0.42$\pm$0.08$^{w21}$ &                   & 2.75$_{-0.18}^{+0.13,w21}$\\
HD\,55606 &               & B2Vnnpe$^{w21}$ & 957$\pm$37 & 0.22$^{w21}$  & 27.35$^{w21}$& 3.52$\pm$0.20$^{w21}$&5.97-6.55$^{c18}$    & 3.60$\pm$0.03 &   93.76$^{c18}$         &75-85$^{c18}$ & 40.9$^{w21}$& 0.27$\pm$0.04$^{w21}$ &0.83-0.9$^{c18}$    & 2.27$_{-0.19}^{+0.13,w21}$\\
HD\,58978 & FY\,CMa       & B0.5IVe$^{w18}$ & 558$\pm$22 & 0.14$^{p08}$  & 27.5$^{p08}$ & 6.8 (5.3-9.7)$^{p08}$&10-13$^{p08}$        & 4.43$\pm$0.03 &   37.255$^{p08}$        &$>$66$^{p08}$ & 45$^{p08}$  & 0.81 (0.6-1.2)$^{p08}$&1.1-1.5$^{p08}$     & 3.38                  \\
HD\,60855 & V378\,Pup     & B3IV$^{w21}$    & 471$\pm$17 & 0.204$^{w21}$ & 20$^{w21}$   & 8.04$\pm$0.62$^{w21}$&                    & 3.99$\pm$0.03 &                        &             & 42$^{w21}$  & 0.49$\pm$0.07$^{w21}$ &                   & 2.83$_{-0.20}^{+0.14,w21}$\\
HD\,113120& LS\,Mus       & B2IVne$^{w21}$  & 424$\pm$15 & 0.176$^{w21}$ & 22.8$^{w21}$ & 5.68$\pm$1.09$^{w21}$&                    & 3.86$\pm$0.03 &                        &             & 45$^{w21}$  & 0.41$\pm$0.14$^{w21}$ &                   & 2.80$_{-0.53}^{+0.23,w21}$\\
HD\,137387& kap01\,Aps    & B2Vnpe$^{w21}$  & 322$\pm$8  & 0.125$^{w21}$ & 23.95$^{w21}$& 5.38$\pm$0.29$^{w21}$&                    & 3.83$\pm$0.02 &                        &             & 40$^{w21}$  & 0.43$\pm$0.06$^{w21}$ &                   & 2.64$_{-0.20}^{+0.14,w21}$\\
HD\,152478& V846\,Ara     & B3Vnpe$^{w21}$  & 298$\pm$5  & 0.252$^{w21}$ & 19.8$^{w21}$ & 4.07$\pm$0.19$^{w21}$&                    & 3.39$\pm$0.01 &                        &             & 42$^{w21}$  & 0.26$\pm$0.04$^{w21}$ &                   & 2.28$_{-0.21}^{+0.14,w21}$\\
HD\,157042& $\iota$\,Ara  & B2.5IVe$^{w21}$ & 279$\pm$7  & 0.187$^{w21}$ & 25.86$^{w21}$& 5.79$\pm$0.42$^{w21}$&                    & 3.95$\pm$0.02 &                        &             & 33.8$^{w21}$& 0.58$\pm$0.09$^{w21}$ &                   & 2.60$_{-0.23}^{+0.15,w21}$\\
HD\,194335& V2119\,Cyg    & B2IIIe$^{w21}$  & 364$\pm$8  & 0.122$^{w21}$ & 25.6$^{w21}$ & 5.13$\pm$0.34$^{w21}$&8.65$\pm$0.35$^{k22}$& 3.83$\pm$0.02 &   63.146$^{k22}$        & 49.4$^{k22}$ & 43.5$^{w21}$& 0.51$\pm$0.07$^{w21}$ &1.62$\pm$0.28$^{k22}$& 2.92$_{-0.23}^{+0.15,w21}$\\
HD\,200120& 59\,Cyg       & B1Ve$^{w18}$    & 414$\pm$59 & 0.041$^s$    & 21.8$^{p13}$ & 6.7$^{p13}$          &6.3-9.4$^{p13}$      & 4.14$\pm$0.12 &  28.1871$^{p13}$        &60-80$^{p13}$ & 52.1$^{p13}$& 0.41$^{p13}$          &0.62-0.91$^{p13}$    & 3.0$\pm$0.1$^{p13}$    \\
HD\,200310& 60\,Cyg       & B1Ve$^{w18}$    & 375$\pm$18 & 0.036$^s$    & 27$^{w17}$   & 5.0$\pm$0.3$^{k22}$  &7.3$\pm$1.1$^{k22}$  & 3.99$\pm$0.04 &   147.68$^{k22}$        & 83.4$^{k22}$ & 42$^{w17}$  & 0.465$^{w17,k22}$      & 1.2$\pm$0.2$^{k22}$ & 2.78                  \\ 
\multicolumn{15}{l}{\it Remaining \citet{wan18} systems}\\
HD\,157832& V750\,Ara     & B1.5Ve$^{w21}$  & 972$\pm$49 & 0.229$^{w21}$ & 25$^{w21}$   &10.71$\pm$0.95$^{w21}$&                    & 4.49$\pm$0.04 &                        &             & 45$^{w21}$  &$<$0.33$^{w21}$       &                     & $<$2.61$^{w21}$         \\
HD\,191610& 28\,Cyg       & B3IVe$^{w21}$   & 255$\pm$7  & 0.182$^{w21}$ & 20.47$^{w21}$& 5.89$\pm$0.37$^{w21}$&                    & 3.76$\pm$0.02 &     246:$^{k22}$        & 118$^{k22}$  & 45$^{w21}$  &$<$0.26$^{w21}$       &                     & $<$2.39$^{w21}$        \\
HD\,214168& 8\,Lac\,A     & B1IVe$^{w21}$   & 520$\pm$27 & 0.025$^{w21}$ & 27.38$^{w21}$& 5.27$\pm$0.98$^{w21}$&                    & 4.17$\pm$0.05 &                        &             & 45$^{w21}$  &$<$0.37$^{w21}$       &                     & $<$2.71$^{w21}$        \\
\multicolumn{15}{l}{\it Other Be binaries}\\
HD\,37202 & $\zeta$\,Tau  & B1IVe          & 136$\pm$6  & 0.044$^{t13}$ & 19.3$^{c19}$ & 6.1$^{c19}$          & 11$^{r09}$          & 3.75$\pm$0.04 & 132.987$^{r09}$         & 60-90$^{r09}$ &            &                    &0.87-1.02$^{r09}$     &                       \\  
          & ALS\,8775     & B3Ve$^{s20}$    &2108$\pm$175& 0.397$^{i20}$ & 18$^{s20}$   & 3.7$^{s20}$          &  7$\pm$2$^{s20}$    & 3.10$\pm$0.07 & 78.7999$^{s20}$         & 39$^{s20}$    & 12.7$^{s20}$&   5.3$^{s20}$        & 1.5$^{s20}$          & 2.8$^{s20}$            \\
HD\,63462 & o\,Pup        & B1IV:nne$^{k12}$& 355$\pm$22 & 0.022$^s$    &             &                     &                    & 4.32$\pm$0.05 &  28.903$^{k12}$         &              &            &                    &0.064 x m(be)$^{k12}$ &                       \\  
HD\,68980 & MX\,Pup       & B1.5III$^{c02}$ & 409$\pm$15 & 0.025$^s$    & 25.1$^{c19}$ & 8.6$^{c19}$          & 15$^{c02}$          & 4.24$\pm$0.03 &  5.1526$^{c02}$         & 5-50$^{c02}$  &            &                    &0.6-6.6$^{c02}$       &                       \\  
HD\,148184& $\chi$\,Oph   & B2Vne          & 153$\pm$4  & 0.354$^{t13}$ & 20.9$^{t08}$ & 5.8$^{t08}$          & 10.9$^{t08}$        & 3.75$\pm$0.02 &34.1$^{h87}$/138.8$^{a78}$& 20$^{h87}$    &           &                      & 3.8$^{h87}$         &                       \\ 
HD\,161306&               & B0:ne$^{k14}$   & 451$\pm$5  & 0.47$^s$     &             &                     &                    & 3.56$\pm$0.01 &    99.9$^{k14}$         &              &            &                    &0.0567 x m(be)$^{k14}$&                       \\ 
HD\,167128& HR6819        & B2.5Ve$^{b20}$  & 368$\pm$17 & 0.089$^s$    & 20$^{b20}$   & 4.2$\pm$0.8$^{b20}$  &  7$\pm$2$^{b20}$    & 3.77$\pm$0.04 &  40.335$^{b20}$         &   32$^{b20}$  & 16$^{b20}$   &  4.8$\pm$0.4$^{b20}$& 0.46$\pm$0.26$^{b20}$& 3.12$\pm$0.10$^{b20}$  \\
\hline
  \end{tabular}
\end{center}

{\footnotesize References: $^{a78}$ \citet{abt78}, $^{b20}$ \citet{bod20}, $^{c02}$ \citet{car02}, $^{c18}$ \citet{cho18}, $^{c19}$ \citep{coc19}, $^{h87}$ \citet{har87}, $^{i20}$ \citet{irr20}, $^{k12}$ \citet{kou12},  $^{k14}$ \citet{kou14}, $^{k22}$ \citet{kle21}, $^{m15}$ \citet{mou15}, $^{p08}$ \citet{pet08}, $^{p13}$ \citet{pet13}, $^{p16}$ \citet{pet16}, $^{r09}$ \citet{rud09}, $^{s20}$ \citet{she20}, $^s$ Stilism (\citealt{lal14} and https://stilism.obspm.fr/), $^{t08}$ \citet{tyc08}, $^{t13}$ \citet{tou13}, $^{w17}$ \citet{wan17}, $^{w21}$ \citep{wan21} and references therein. When no reference for spectral type is provided, that coming from Simbad is used. If the distance used in the radius reference is different from the one adopted here, a scaling was made. Note that the errors for bolometric luminosities reflect only the errors on the distances.}

\end{table}
\end{landscape}

Massive stars often lie in binary systems and \gc\ are no exception. Indeed, two \gc\ stars have long been known to lie in a binary system: \gc\ \citep{har00} and $\pi$\,Aqr \citep{bjo02}. The multiplicity of other \gc\ analogs has been recently studied \citep{naz21}. In total, orbits could be derived in eight cases (out of 25 \gc\ objects) and five additional stars show hints of binarity. The remaining Be stars were either too faint or had too variable line profiles to study binarity signatures. The observed properties of the \gc\ analogs (long periods and small velocity amplitudes) are similar to those seen in other Be binaries and point towards companions with low masses. However, their exact nature could not yet be determined: white dwarfs, neutron stars, late-type non-degenerate stars, and stripped stars have all been proposed at some point.

Even though the presence of stripped stars in some \gc\ analogs remains circumstantial, there are several known cases of Be stars with a stripped-star companion (called Be+sdO or BeHeB). Such companions were detected thanks to their high temperature, which gives them an apparent spectral type O. The high temperature also leads to a strong UV emission. The first cases were reported in analyses of individual systems: HD\,10516 ($\phi$\,Per, \citealt{gie98,mou15}), HD\,58978 (FY\,CMa, \citealt{pet08}), HD\,41335 (HR2142, \citealt{pet16}), HD\,200120 (59\,Cyg, \citealt{pet13}), and HD\,200310 (60\,Cyg, \citealt{kou00,wan17}). Then a general search in IUE data detected twelve additional cases \citep{wan18}. Follow-up HST spectroscopy confirmed the presence of a stripped companion to nine of these, with an additional case not previously studied \citep{wan21}. The three unconfirmed cases would be analogs to HD\,41335, where the signature of the stripped star was detected only at some orbital phases. This suggests that the stripped companion does exist, but is less luminous and cooler than in other systems. 

The X-ray properties of these 18 Be+sdO systems are largely unknown since they were not studied in detail. With two exceptions (see below), only flux limits are reported for some cases in the literature \citep{ber96}. This paper thus tries to fill this gap by performing a global X-ray study of these Be+sdO systems, with the ultimate aim of evaluating the validity of the Langer et al. scenario. Additional Be binaries, for which the nature of the companion is unknown but could be a stripped star \citep[e.g. HD\,161306, ][]{kou14}, were also added to the sample for completeness. Section 2 presents the collected data while Section 3 reports on their analysis. Section 4 then discusses the Langer et al. scenario, with the constraints brought by previous data and our observations highlighted. Section 5 finally summarizes our findings.

\section{Data}

\subsection{The sample}

Our sample consists of the 18 known Be+sdO systems plus seven spectroscopic Be binaries never studied in X-rays and for which the nature of the companion remains unclear. Table \ref{opt} lists those stars, along with their known properties. The distances were taken from \citet{bai21}, except for HD\,37202 (which has no {\it Gaia} data, and so the {\it Hipparcos} value of \citealt{van07} is used) and HD\,200120 (which has a clearly deviant distance in the {\it Gaia} catalog, \citealt{pet13,naz18}). When this distance was different from that adopted in the references used for stellar radii of the stars and for bolometric luminosities of the companions, these parameters were appropriately scaled. Bolometric luminosities of the Be stars were estimated using the $V$-band magnitudes from Simbad, the interstellar reddenings and distances quoted in the Table, as well as bolometric corrections derived for the adopted temperatures of the Be stars using the formula of \citet{ped20}. When the $T_{eff}$(Be) was unknown, the correction used for stars with the closest spectral types were used. We find that these bolometric luminosities agree well with those derived using the temperatures and the radii, when known. 

\subsection{X-ray observations}

Most of our targets were observed in 2021 with \xmm\ in the framework of our dedicated program \#088003. In addition, HD\,200120 had been previously observed for us during another project \citep{naz20} and two additional archival datasets of GJ\,674 (0551020101, PI Schmitt - see also \citealt{naz18}, and 0810210301, PI Froning) observed HD\,157832 off-axis. All \xmm\ data were processed with the Science Analysis Software (SAS) v20.0.0 using calibration files available in January 2022 and following the recommendations of the \xmm\ team\footnote{SAS threads, see \\ http://xmm.esac.esa.int/sas/current/documentation/threads/ }. After their pipeline processing, the European Photon Imaging Camera (EPIC) event files were filtered to keep only the best-quality data ({\sc{pattern}} 0--12 for MOS and 0--4 for pn). Light curves for energies above 10\,keV revealed contamination by background proton flares for all but the HD\,51354, HD\,152478, and HD\,200120 datasets. Thresholds on lightcurve count rates above 10\,keV were then applied to eliminate the flaring intervals.

Source detection was performed in the total, 0.5--10.\,keV, energy band, as well as for soft (0.5--2.\,keV) and hard (2.--10.\,keV) bands to constrain the hardness of the detected X-ray sources. Images binned by a factor of 20, i.e. to a 1\arcsec\ pixel size, and likelihoods of 10 were used. A trial was also performed with lower thresholds. This resulted in the detection of two additional sources, associated to HD\,137387 and HD\,157042. Both have combined EPIC likelihoods $\sim$10, which was our initial limit, and so we have added them to the list of detections. For HD\,152478, an X-ray source is detected 18.7\arcsec\ away from the Simbad J2000 position or {\it Gaia} J2016 position of the target. This difference is much too large for a secure association \citep[its positional error is 0.8\arcsec\ and the absolute measurement accuracy of \xmm\ is 4\arcsec\ half-cone angle, see][]{jan01} and we therefore tag this source as undetected. Sensitivity maps using the standard Poisson mode were built for a likelihood of 3.0 (corresponding to a probability of 95\%), to estimate upper limits on count rates for undetected targets. Table \ref{donx} provides the final count rates or upper limits on count rates. For completeness, we may note that an alternative way to build such maps exists in {\sc SAS}. It relies on the ``delta-C'' mode but requires larger pixels to be applied. Using 4\arcsec\ pixels, this method leads to limits 30 to 100\% larger than those of the standard mode.

When a target was bright enough, \xmm\ spectra were extracted for each EPIC camera. The source regions were circles centered on the Simbad positions of the targets and with typical radii of 30\arcsec\ while background regions were chosen from nearby circles devoid of sources. For HD\,194335 and HD\,200120, the individual EPIC spectra have low signal-to-noise hence we combined them using the task {\sc epicspeccombine}. A grouping was applied to all \xmm\ spectra to obtain an oversampling factor of maximum five and a minimum signal-to-noise ratio of 3.

Four additional targets were covered by \ch\ observations, two for our dedicated program (ObsIDs 25112, 25115, 25116) and two from archival datasets (ObsIDs 20928, PI Liu, and 26239, PI Koss). Source detection was performed in the total, 0.5--10.\,keV, energy band using both a sliding square detect cell (task {\sc celldetect}) and wavelets (task {\sc wavdetect}). This led to two detections: HD\,37202 and HD\,58978. In the former case, the detection occurs at 1.3\arcsec\ of the Simbad position of the source (J2000) or 0.9\arcsec\ of the {\it Gaia} position (J2016). This is quite far away for \ch, but not unheard of as ``the 99\% limit on positional accuracy is 1.4\arcsec'' for \ch\footnote{https://cxc.harvard.edu/mta/ASPECT/celmon/}. Also, the target appears quite isolated, with no known star or extragalactic source nearby. Therefore we consider the detection as secure. However, this target appears extremely bright in X-rays ($>0.2$\,cts\,s$^{-1}$) and is therefore affected by pile-up. Its spectrum was derived using the task {\sc specextract}, a source region of 20 pixels radius centered on the {\sc wavdetect} position and a surrounding annular background region of 20 and 70 pixels radii. The examination of the spectrum clearly reveal an extremely hard tail typical of pile-up. The use of the ``pileup'' model within Xspec did not lead to a good result. Therefore, a new extraction was performed, using an annular source region with 3 and 20 pixels radii. The inner radius was chosen as a compromise between pile-up elimination and sensitivity, after trials with 1, 3, and 5 pixels for inner radius. Such annular extractions clearly showed the progressive disappearance of the tail, confirming its origin as due to pile-up. However, to obtain the correct effective area for the annular source region, the arf had to be modified using the task {\sc arfcorr}. A grouping by minimum 10 counts led to the final spectra. 

For HD\,58978, the wavelet detection algorithm found it but with only $\sim$6 counts (all appearing in the soft X-ray range). For this target as well as the two undetected ones, the \ch\ count rates were estimated in the 0.5--10.\,keV energy band using the task {\sc srcflux} for a source region of 10 pixels radius centered on the Simbad positions of the targets and an annular background region of 10 and 50 pixels radii. The confidence level was set to 68\% to find the 1$\sigma$ error on the rate of HD\,58978 while a level of 95\% was used to derive upper limits (Table \ref{donx}). Note that the listed value for HD\,41335 corresponds to the combination of two observations, which improves the limit determination.

Some observations from the Neil Gehrels \sw\ observatory covered our targets and were processed with the on-line tool\footnote{https://www.swift.ac.uk/user\_objects/}. However, the \sw\ X-ray telescope (XRT) suffers from optical loading in case of optically-bright sources. This restricted the use of \sw\ data taken in PC mode to HD\,55606 and ALS\,8775. For the latter, a stricter limit on its X-ray emission was provided by the \ch\ data and so the \sw\ data were not used. In parallel, we requested an observation of HD\,10516 in WT mode, to avoid optical loading. The derived 2$\sigma$ upper limits on the sources' count rates (in the 0.3--10.\,keV band) are provided in Table \ref{donx}.

Finally, the remaining targets were not observed during dedicated pointed observations but were sampled during \xmm\ slews between observations. The upper limit server\footnote{http://xmmuls.esac.esa.int/upperlimitserver/} (hereafter ``uls'') provided in such cases upper limits on pn count rates in the 0.2--12.\,keV energy band and with 2$\sigma$ significance (corresponding to 95\% probability). These values are listed in Table \ref{donx}. 

\begin{table*}
  \scriptsize
  \caption{Observed X-ray characteristics of the targets. \label{donx}}
  \begin{tabular}{lcccccccccccc}
    \hline
Name & ObsID & $HJD$ & $N_{\rm H}^{ISM}$ & \multicolumn{3}{c}{Count rates ($10^{-3}$\,cts\,s$^{-1}$)} & \multicolumn{3}{c}{Count hardness} &$L_{\rm X}^{ISM-cor}$ & $\log(L_{\rm X}/L_{\rm BOL})$ & $EW$(H$\alpha$) \\
     &       & $-2.45e6$& ($10^{22}$   &      & XRT/ACIS/& & \multicolumn{3}{c}{} & (tot,  &  & (\AA)\\
     &       &          & \,cm$^{-2}$) & MOS1 & MOS2 & pn  & MOS1 & MOS2 & pn & $10^{30}$\,erg\,s$^{-1}$)& \\
\hline                                               
\multicolumn{11}{l}{\it Bright detections}\\
HD\,194335& 0880030501$^x$ & 9364.813& 0.075  & 4.80$\pm$0.85 & 4.54$\pm$0.90 & 19.0$\pm$2.1 & $-0.72\pm$0.14 & $-0.82\pm$0.14 & $-0.89\pm$0.06 & 0.55$\pm$0.03 & $-7.67\pm$0.02    & --1.3 \\
HD\,200120& 0820310501$^x$ & 8255.971& 0.025  & 4.34$\pm$0.87 & 3.87$\pm$0.74 & 32.4$\pm$2.4 & $-1.00\pm$0.04 & $-0.94\pm$0.08 & $-0.99\pm$0.03 & 0.84$\pm$0.27 & $-7.80\pm$0.07    &--11.1 \\ 
HD\,157832& 0551020101$^x$ & 4715.251& 0.14   & 244$\pm$6     & 264$\pm$7     & 631$\pm$11   & $-0.09\pm$0.02 & 0.08$\pm$0.03  & $-0.15\pm$0.02 & 320$\pm$33    &$-5.568\pm$0.008   &       \\
HD\,157832& 0810210301$^x$ & 8211.944& 0.14   & 172$\pm$5     & 243$\pm$7     &              & 0.14$\pm$0.03  & 0.35$\pm$0.03  &                & 285$\pm$30    &$-5.618\pm$0.011   &--28.0 \\
HD\,37202 & 26239$^c$      & 9573.664& 0.027  &               &               &              &                &                &                & 36.7$\pm$9.2  & $-5.77\pm$0.05    &--13.4 \\
\multicolumn{11}{l}{\it Faint detections}\\                                
HD\,43544 & 0880031301$^x$ & 9498.389& 0.095  & 1.07$\pm$0.50 & 0.85$\pm$0.49 & 3.00$\pm$1.07& $-1.00\pm$0.11 & $-1.00\pm$0.18 & $-1.00\pm$0.20 & 0.11$\pm$0.03 & $-8.08\pm$0.01      &--18.6 \\
HD\,58978 & 25112$^c$      & 9557.777& 0.086  & \multicolumn{3}{c}{1.10 (0.64..1.7)}         & \multicolumn{3}{c}{}                             & 4.1 (2.4..6.3)& $-7.4\,(-7.6..-7.2)$&       \\
HD\,137387& 0880030201$^x$ & 9442.584& 0.077  & 0.87$\pm$0.51 & 0.66$\pm$0.45 & 2.91$\pm$1.12& $-0.11\pm$0.57 & $-1.00\pm$0.27 & $-1.00\pm$0.20 & 0.09$\pm$0.03 & $-8.46\pm$0.01      &       \\
HD\,157042& 0880030701$^x$ & 9468.515& 0.11   & 1.02$\pm$0.66 & 1.06$\pm$0.52 & 3.61$\pm$1.19& $-1.00\pm$0.22 & $-0.90\pm$0.31 & $-0.93\pm$0.23 & 0.09$\pm$0.02 & $-8.56\pm$0.01      &       \\
HD\,191610& 0880031101$^x$ & 9526.353& 0.11   & 1.23$\pm$0.64 & 1.87$\pm$0.68 & 5.15$\pm$1.52& $-0.87\pm$0.27 & $-0.67\pm$0.32 & $-1.00\pm$0.26 & 0.15$\pm$0.04 & $-8.17\pm$0.01      & --8.8 \\
\multicolumn{8}{l}{\it Non-detections}\\
HD\,10516 & 00015011001$^s$& 9595.940& 0.099  & \multicolumn{3}{c}{$<$15.} & & & &  $<$1.6..2.6 & $<-7.5..-7.3$ &--33.0 \\
HD\,29441 & 0880031501$^x$ & 9632.720& 0.12   & $<$0.32 & $<$0.33 & $<$0.35& & & & $<$0.03..0.17& $<-8.6..-7.8$ &--28.9 \\
HD\,41335 & 25115$^c$      & 9711.340& 0.061  & \multicolumn{3}{c}{$<$0.57}& & & &  $<$0.31..1.7& $<-8.3..-7.5$ &       \\
          & 25116$^c$      & 9568.596&                                                                                  \\
HD\,51354 & 0880031401$^x$ & 9660.617& 0.11   & $<$0.41 & $<$0.41 & $<$0.48& & & & $<$0.03..0.16& $<-8.6..-7.9$ &       \\
HD\,55606 & 00011412001$^s$& 8627.005& 0.13   & \multicolumn{3}{c}{$<$3.6} & & & &  $<$13..19   & $<-6.1..-5.9$ &       \\
HD\,60855 & 0880030301$^x$ & 9508.014& 0.12   & $<$0.92 & $<$1.1  & $<$2.8 & & & & $<$0.17..0.42& $<-8.3..-7.9$ &       \\
HD\,113120& 0880030601$^x$ & 9444.580& 0.11   & $<$1.0  & $<$1.1  & $<$2.1 & & & & $<$0.10..0.30& $<-8.4..-8.0$ &       \\
HD\,152478& 0880030801$^x$ & 9626.659& 0.15   & $<$0.83 & $<$0.84 & $<$1.4 & & & & $<$0.03..0.13& $<-8.5..-7.8$ &       \\
HD\,200310& uls$^x$        & 3484.186& 0.022  &         &         & $<$690 & & & &  $<$9.1..30  & $<-6.6..-6.1$ &       \\
HD\,214168& 0880031001$^x$ & 9408.685& 0.015  & $<$0.47 & $<$0.50 & $<$0.99& & & & $<$0.05..0.18& $<-9.1..-8.5$ & 3.8   \\
ALS\,8775 & 20928$^c$      & 8132.204& 0.24   & \multicolumn{3}{c}{$<$0.96}& & & &  $<$10..80   & $<-5.7..-4.8$ &       \\
HD\,63462 & uls$^x$        & 5876.503& 0.013  &         &         & $<$470 & & & &  $<$5.0..18  & $<-7.2..-6.6$ &       \\
HD\,68980 & uls$^x$        & 9146.051& 0.015  &         &         & $<$390 & & & &  $<$56..194  & $<-6.1..-5.5$ &       \\
HD\,148184& uls$^x$        & 6905.410& 0.22   &         &         & $<$830 & & & &  $<$7.8..10  & $<-6.4..-6.3$ &--41.2 \\
HD\,161306& uls$^x$        & 3999.213& 0.29   &         &         & $<$100 & & & &  $<$7.3..11  & $<-6.3..-6.1$ &       \\
HD\,167128& uls$^x$        & 3648.710& 0.054  &         &         & $<$610 & & & &  $<$11..29   & $<-6.3..-5.9$ &       \\
\hline
  \end{tabular}
  
{\scriptsize A superscript on the ObsID identifies the facility used (x for \xmm, c for {\it Chandra}, s for {\it Swift}). The interstellar hydrogen columns were evaluated using the reddening from Table \ref{opt} and the relation $N_{\rm H}^{ISM}=6.12\times 10^{21}\times E(B-V)$\,cm$^{-2}$ \citep{gud12}. For \xmm, the count rates are provided for pn in the 0.2--12.\,keV energy band for the uls data, and for MOS1, MOS2, and pn in the 0.5--10.\,keV energy band otherwise. The associated count hardnesses are calculated from $(cr_h-cr_s)/(cr_h+cr_s)$ where $cr_h$ and $cr_s$ correspond to count rates in the energy bands 2.--10.\,keV and 0.5--2.\,keV, respectively. For \ch, count rates are for ACIS-S in the 0.5--10.\,keV energy band. For \sw, count rates are for XRT in the 0.3--10.\,keV energy band. All upper limits are 95\%, and all X-ray luminosities $L_{\rm X}^{ISM-cor}$ provided after correction for interstellar absorption and in the 0.5--10.\,keV energy band. The last column provides equivalent widths $EW$ of the H$\alpha$ line, evaluated from --600\,km\,s$^{-1}$ to 600\,km\,s$^{-1}$ in the closest optical spectrum available in the Bess database (\citealt{nei11}, http://basebe.obspm.fr/basebe/). The spectra were considered only if taken within six months of the X-ray dataset (but most often are taken within weeks).}
\end{table*}

\begin{table*}
  \caption{Best-fit models to the X-ray spectra. \label{fitsx}}
  \begin{tabular}{lccccccc}
    \hline
Name & $N_{\rm H}$ & $kT$ & $norm$ & $\chi^2$/dof & $F_{\rm X}^{obs}$ & $F_{\rm X}^{ISM-cor}$ & $HR$ \\
     &($10^{22}$\,cm$^{-2}$) & (keV) & (cm$^{-5}$) & & \multicolumn{2}{c}{(tot, erg\,cm$^{-2}$\,s$^{-1}$)} & \\
\hline                                               
HD\,194335    & 0.96$\pm$0.12 &0.15$\pm$0.02 & (4.11$\pm$6.71)e-3 &28.89/16   & (2.79$\pm$0.10)e-14  &3.48e-14  &(7.5$\pm$0.8)e-4\\
HD\,200120    & 0.00$\pm$0.01 &0.27$\pm$0.01 & (3.71$\pm$0.36)e-5 &38.23/25   & (3.74$\pm$0.57)e-14  &4.10e-14  &(1.2$\pm$0.4)e-3\\
HD\,157832$^a$& 0.14$\pm$0.01 &9.43$\pm$0.79 & (1.65$\pm$0.02)e-3 &429.07/413 & (2.68$\pm$0.05)e-12  &2.83e-12  &2.97$\pm$0.08\\
HD\,157832$^b$& 0.63$\pm$0.05 &9.43 (fixed)  & (1.70$\pm$0.05)e-3 &192.45/200 & (2.46$\pm$0.06)e-12  &2.52e-12  &5.10$\pm$0.20 \\
HD\,37202     & 15.5$\pm$2.0  &11.1$\pm$4.3  & (2.29$\pm$0.26)e-2 &30.01/38   & (1.66$\pm$0.20)e-11  &1.66e-11  &2392$\pm$1831\\
\hline
  \end{tabular}
  
{\scriptsize Fitted models were of the form phabs$\times$phabs$\times$apec, with the first absorption fixed to the interstellar value (see Table \ref{donx}). $^{a,b}$ refer to \xmm\ datasets with ObsID 0551020101 and 0810210301, respectively. The flux hardness ratios are defined by $HR = F_{\rm X}^{ISM-cor}(hard)/F_{\rm X}^{ISM-cor}(soft)$, with $F_{\rm X}^{ISM-cor}$ the flux after correction for interstellar absorption and soft and hard energy bands being defined as 0.5--2.0 keV and 2.0--10.0 keV, respectively (the total band being 0.5--10.0 keV). Errors correspond to 1$\sigma$ uncertainties; they correspond to the larger value if the error bar is asymmetric. }
\end{table*}

\section{Results}

X-ray observations of four targets collected enough counts for a meaningful extraction of spectra. These spectra were then fitted in {\sc Xspec} v12.11.1 using absorbed optically thin thermal emission models with solar abundances from \citet{asp09}. The chosen models were similar to those of \citet{naz18} and results are provided in Table \ref{fitsx}. For \xmm, note that all EPIC spectra (pn, MOS1, MOS2) of each exposure of HD\,157832 were fitted simultaneously. Moreover, as the fitting to the second exposure leads to an extremely high temperature which cannot be constrained ($kT>19$\,keV), an alternative fitting fixing the temperature to that of the first exposure was tried. As both fitting attempts yield similar reduced $\chi^2_{\nu}$ (just below 1), only the alternative results, with lower flux errors, were kept. Slight differences with results reported in \citet{naz18} and \citet{naz20} for the same exposures arise from a combination of small analysis changes (improved atomic parameters of the fitting tool, combination of spectra for HD\,200120, slightly lower distance and stronger interstellar absorption for HD\,157832). They do not alter the conclusions in this or these previous papers. 

\begin{figure}
  \begin{center}
    \includegraphics[width=8cm]{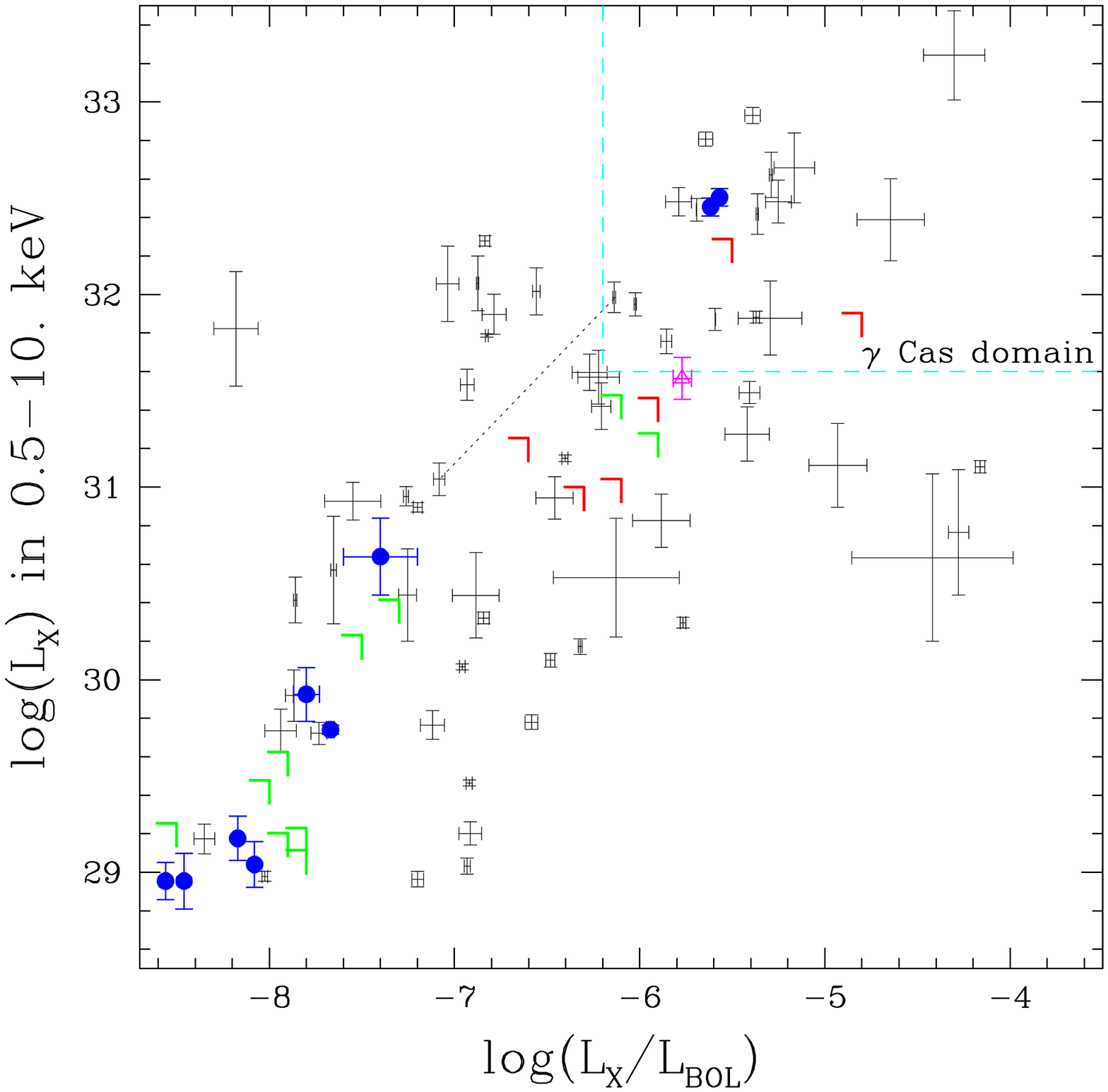}
    \includegraphics[width=8cm]{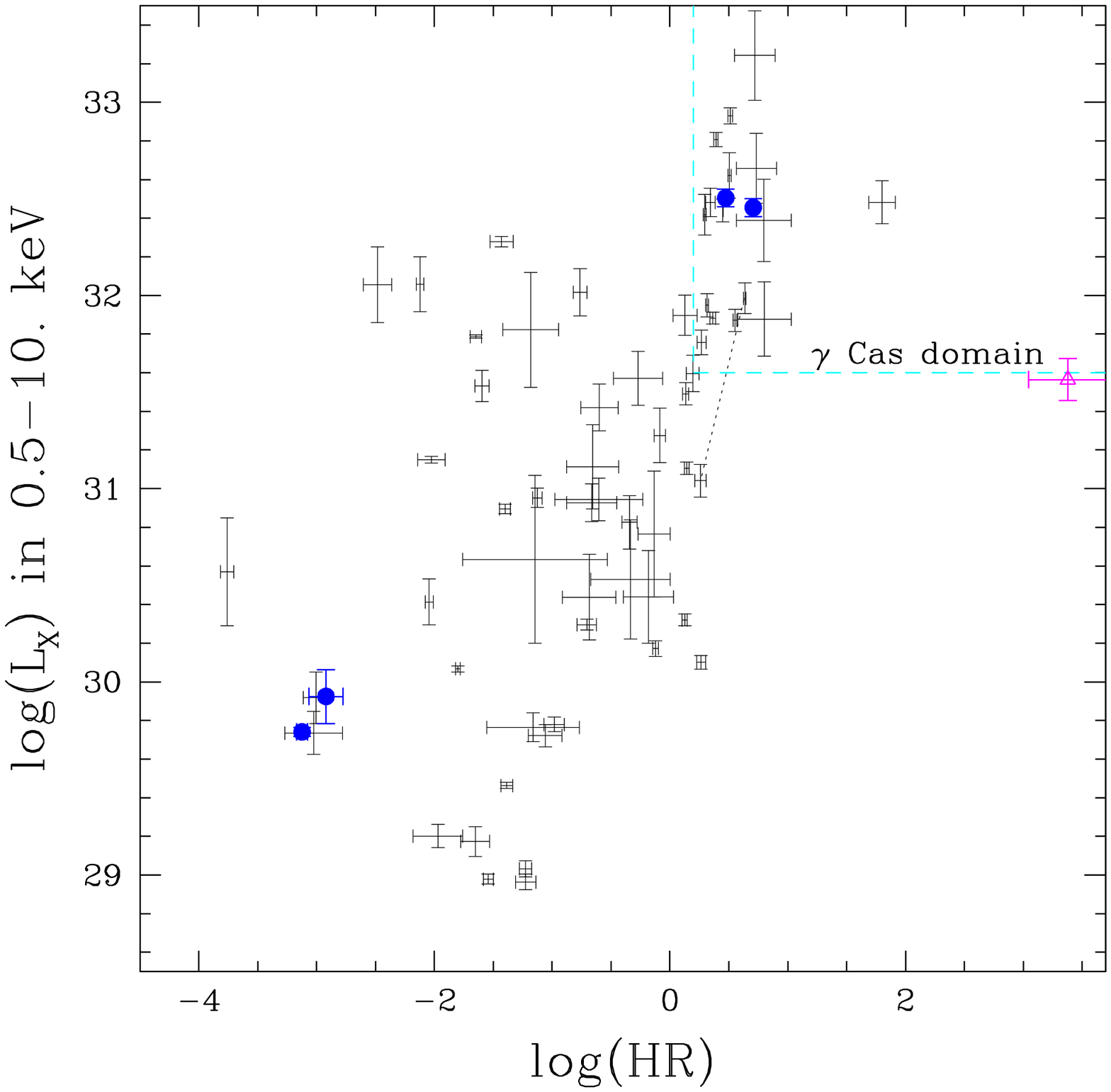}
  \end{center}
  \caption{Comparison of hardness $HR$ and brightness $L_{\rm X}$ for our sample of Be systems and those presented in \citet{naz18} and \citet{naz20}. Blue circles are used for Be+sdO systems with detected X-ray emissions, green symbols for the 95\% upper limits of undetected Be+sdO systems, a magenta triangle for the Be binary with detected X-rays, red symbols for the 95\% upper limits of the undetected Be binaries, and black symbols for objects reported in previous studies. Note that HD\,157832 is represented twice as two \xmm\ observations are available.}
\label{lx}
\end{figure}

For five targets faintly detected in X-rays, only a count rate is available. In such cases, the count rates need to be converted to obtain X-ray fluxes, which was done with the WebPIMMS tool\footnote{https://heasarc.gsfc.nasa.gov/cgi-bin/Tools/w3pimms/w3pimms.pl}. To this aim, we have assumed an absorbed optically thin thermal emission model, with its absorption fixed to the interstellar one and a temperature of 0.2\,keV. This temperature was chosen because all sources appear soft. For example, in \xmm\ data, the proportions of soft and hard counts in HD\,43544, HD\,137387, HD\,157042 and HD\,191610 agree with those of HD\,194335 and HD\,200120 (see columns count hardnesses in Table \ref{donx}) and the spectral fits of the latter two sources suggest a low plasma temperature (Table \ref{fitsx}). The derived luminosities are presented in Table \ref{donx}.

For the 16 undetected targets, only a 95\% upper limit on the count rates could be derived. Again, they were converted using WebPIMMS and an absorbed thermal model. The absorption was again fixed to the interstellar one but, in the absence of any hint of the hardness, three temperatures ($kT$=0.2, 0.6, and 5.4\,keV or $\log(T)$=6.4, 6.85, and 7.8) were tried. This led to a range of X-ray luminosities (Table \ref{donx}).

With the X-ray properties at hand for all systems, we can now examine the results. First of all, a comparison with the known properties of Be stars in the X-ray range \citep{naz18,naz20} should be made. Figure \ref{lx} compares the brightness and hardness of our targets to those of other Be samples. Here, brightness corresponds to the X-ray luminosity $L_{\rm X}$ in the 0.5--10.\,keV energy band, while the hardness $HR$ is the ratio of hard (2.--10.\,keV) to soft (0.5--2.\,keV) X-ray fluxes after correction for interstellar absorption. Neither the Be+sdO systems nor the other Be binaries seem to display X-ray characteristics differing from those reported in previous Be X-ray studies. The known binarity of our targets thus does not seem to have a significant impact overall on the observed X-ray properties of Be stars. Incidentally, this result may appear quite normal if one considers that most (if not all) Be stars are binaries, often with a stripped-star companion \citep[e.g.][]{sha14,kle19}. Nonetheless, one could argue that the detected companions of our targets represent the tip of the iceberg in this context. They would therefore be special in some way (e.g. hotter, more massive) and the absence of consequences on the X-ray emission then remains a significant result.

  In parallel, one may be tempted to argue against the presence of a massive compact companion when X-ray emission well below the X-ray binary luminosity range is detected for a Be star. However, caution should be applied here, as such X-ray luminosities can only discard the presence of ``usual'', actively accreting objects. Indeed, some neutron stars may appear X-ray fainter if in the propeller regime \citep{pos17} and some black holes may be quiescent hence X-ray faint \citep{psz08,rey14}. It may finally be noted that the presence of black hole companions to our targets ALS\,8775 and HD\,167128 (see references in Table \ref{opt}), as well as to the Be star HD\,215227 (=MWC\,656, see \citealt{ale15} and Rivinius et al., in press), has been recently rejected. Their high-energy characteristics are thus not peculiar anymore.

Second, we may wonder whether some of these stars are \gc\ in character. \citet{naz18} and \citet{naz20} listed the criteria for such a classification. Two of them are particularly relevant for this study (others require more detailed information): brightness of the X-ray emission ($\log(L_{\rm X}/L_{\rm BOL})$ between --6.2 and --4, $\log(L_{\rm X})$ between 31.6 and 33.2 for the total band) and hardness of the X-ray emission ($HR>1.6$, $kT>$5\,keV, $\log(L_{\rm X})>31$ for the hard band). Using these criteria, one star of the sample appears to be a \gc\ analog: the Be+sdO system HD\,157832. However, this has already been known for a decade \citep{lop11}. One more exposure is now available, compared to this discovery paper, and it reveals some variations in flux and hardness, as often seen in \gc\ stars. However, it must be noted that the star fulfills the \gc\ criteria at both epochs. One additional target displays interesting properties: the Be binary HD\,37202 ($\zeta$\,Tau). It fulfills the $\log(L_{\rm X}/L_{\rm BOL})$, $HR$, $\log(L^{hard}_{\rm X})$ criteria, but is {\it exactly} at the limit for the 0.5--10.\,keV X-ray luminosity ($\log(L^{tot}_{\rm X})=31.6$). In this context, it needs to be noted that HD\,37202 appears as an extreme case. Indeed, its spectrum shows few counts below 2\,keV and that is due to a very strong circumstellar absorption ($1.6\times10^{23}$\,cm$^{-2}$, Table \ref{fitsx}). This lowers the X-ray luminosity in the full, 0.5--10.\,keV, energy band. Therefore, it appears quite logical to add HD\,37202 to the list of \gc\ stars. One possible avenue to explain the high absorption of this target could be its high inclination as it is well known that its Be disk is seen edge-on \citep{qui97,gie07,car09,tou11}. Only a few \gc\ analogs were previously found to have strong X-ray absorptions, i.e. $N_{\rm H}\gtrsim10^{22}$\,cm$^{-2}$ \citep{naz18,naz20}, although not as extreme as HD\,37202: V782\,Cas, V771\,Sgr, and V2156\,Cyg. Unfortunately, no evaluation of the disk inclination can be found for them in the literature. However, it may be noted that other \gc\ objects, for which such evaluation exists, have lower inclinations {\it and} do not display high absorbing columns. Trying to put the X-ray data in context, we have also analyzed the recent optical data of HD\,37202, which reveal large differences with previously reported behaviour (see Appendix).

Regarding other stars, the quite high upper limits derived for two stars (ALS\,8775, HD\,68980) do not permit to formally exclude a \gc\ nature for them (Fig. \ref{lx}). Six additional ones (HD\,55606, HD\,63462, HD\,148184, HD\,161306, HD\,167128, HD\,200310) have limits close, albeit clearly lower, than the \gc\ threshold. It would therefore be surprising if they were \gc\ analogs (Fig. \ref{lx}). For HD\,167128, \citet{man21} reported an X-ray detection in the soft X-ray range by {\it AstroSat}. The quoted luminosity ($5\times10^{-12}$\,erg\,cm$^{-2}$\,s$^{-1}$ in 0.3--3.0\,keV, no error mentioned) and quoted spectral properties ($\sim$0.12\,keV blackbody temperature) would lead to $L_{\rm X}\sim5\times10^{31}$\,erg\,s in the 0.5--10.\,keV band, or $\log(L_{\rm X}/L_{\rm BOL})\sim-5.6$), just slightly above our upper limit. Details are very scarce in this Telegram, but the softness of the X-ray emission clearly excludes the target from the \gc\ category. If the X-ray luminosity level is confirmed, HD\,167128 would be somewhat atypical amongst Be stars, the soft X-ray cases being generally fainter. The last eight targets with limits on their X-ray emission as well as the seven other detected sources are far from the \gc\ domain. Therefore, they certainly do not belong to the \gc\ category, without any ambiguity (Fig. \ref{lx}). Their X-ray emission is clearly faint and, when known, soft.

There are no comprehensive and unbiased statistics yet available enabling us to derive a secure incidence rate of \gc\ stars amongst Be stars - only hints can be derived. Nevertheless, if we compare the three large-scale Be studies in the X-ray range, we find that the fraction of \gc\ stars was 15/84 in \citet{naz18}, 3/18 in \citet{naz20}, and 2/25 (one amongst 18 Be+sdO and one amongst 7 other Be binaries) in this work. Alternatively, 66/84, 14/18, and 21/25 of the Be stars in the same papers were definitely not \gc\ or candidates. Despite the small number statistics and the different biases\footnote{This work examines only known binaries, \citet{naz20} studied some stars which could a priori be interesting candidates for presenting the \gc\ phenomenon, and \citet{naz18} analyzed all available X-ray exposures covering Be stars in the last two decades - some data having been required specifically to study \gc\ stars while others serendipitously covered the position of a Be star in the field-of-view. It may be also noted that both studies mostly considered early-type ($<$B3) Be stars, which also applies to the targets of this paper. }, these fractions appear remarkably similar, with about 80\% of non-\gc\ stars and 10--20\% of \gc\ stars in these samples. With current data, it thus appears that our sample of Be binaries is fully in line with the known Be population. Combining the three samples, we then find a global incidence rate of the \gc\ phenomenon of 19/125$\sim$15\% amongst Be stars observed in the X-ray range.

\section{Discussion}
\citet{gie98} and \citet{kri16} have independently imagined the possibility of a collision between the Be disk material and the wind of a stripped star companion. \citet{lan20} hearkened back this idea and applied it with the aim of reproducing the \gc\ peculiarities. In this context, we will examine each step of their scenario in turn, beginning with the properties of stripped stars, then looking into the characteristics of colliding-wind systems and of additional features presented as supportive of their scenario, to conclude with the constraints brought by the actual X-ray data presented in the previous section. 

\subsection{The properties of stripped stars}
By definition, hot subluminous stars, classified using the spectral types sdO, display luminosities below those of the corresponding main-sequence stars (for a review, see e.g. \citealt{heb16}). They are considered as stellar cores stripped of most of their envelopes and burning helium in the core or in a shell around it (hence the alternative name of ``Helium-stars''). The common features of all such stars are their low masses and relatively high effective temperatures. Regarding the origin of these objects, evolutionary effects appear to play a prominent role. However, no single evolutionary path can be identified. The stripping may thus occur through a helium flash after leaving the red giant branch, a merger event, or an interaction between stars in a binary system. 

The He-star companions to Be stars are expected to come from the latter channel. Detailed models of such systems were made by several authors \citep[e.g. the recents studies of][]{sha14,sha21}. \citet{sha14} evolved a population of massive systems, considering a range of periods, masses, and mass-transfer properties. They found that the observed mass distribution of Be stars having a neutron star companion and ranging from 8 to 22\,M$_{\odot}$ is best reproduced in the case of moderately non-conservative mass-transfer without any common envelope phase. In addition, the vast majority of Be binaries should possess white dwarf or He-star companions, neutron star or black holes being orders of magnitude less frequent. \citet{sha21} further showed that their models reproduce well the properties of the observed Be+sdO systems. They also revealed a correlation between the Be star and the He-star masses, with few (if any) stripped stars above 4\,M$_{\odot}$. 

As they emit most of their light in the ultraviolet range, it seems quite natural to expect line-driven outflows for these stripped stars. Models of massive stars were extended (or just applied) by some authors to study these objects, while several pieces of observational evidence revealed direct signature of their winds. 

Theoretically, atmosphere models of stripped stars were made by several authors. \citet{kri16} limit models to a single mass (0.5\,M$_{\odot}$) with a range of temperatures and radii. They demonstrated that the winds may be difficult to launch for stars with high gravities. Also, with too low densities, ion decoupling may occur, which strongly limits the acceleration of the bulk material. Finally, they found ratios between wind velocities and escape velocities to typically lie between 1.5 and 2.5, i.e. lower than for the winds of O-stars. \citet{vin17} rather considers objects with a single effective temperature (50\,kK) and a range of masses. In those models, the mass-loss rates and wind velocities have higher values than in \citet{kri16}, although wind velocities remain below 3000\,\kms\ for stripped stars with masses $<4$\,M$_{\odot}$ (the useful range as it corresponds to the observed objects, see below). For these stars, the wind-to-escape velocity ratio can be derived to be 1.7--3.1. Finally, \citet{got18} studied the evolution of various binaries, leading to stripped stars with a wide range of masses. They do not determine the wind properties, but use the mass-loss formula of \citet{kri16} and a wind-to-escape velocity ratio of 1.5. It may be underlined that their low-mass models reproduce well the observed properties of stripped stars in Be+sdO systems \citep{wan21}.

Observationally, there is little doubt that stripped stars may power winds. First, since those line-driven winds should be as unstable as those of OB stars, shocks should occur and X-ray emission similar to that recorded for single, non-magnetic, OB stars could be expected. This was indeed detected in a handful cases of sdO stars, although all sdBs and most sdOs remained undetected in the X-ray range, most probably because of too tenuous winds \citep[for a review, see][]{mer16}. These X-ray emissions displayed plasma temperatures and bolometric-to-X-ray luminosity ratios similar to those of classical Population I OB stars, a consistency check of their common origin. 

In addition, some subdwarf spectra display lines clearly associated with outflows, notably P Cygni profiles in some UV lines. Atmosphere modelling was then performed to determine the stellar and wind properties. \citet{jef10} fitted the spectra of six putatively isolated ``extreme He-stars'', deriving effective temperatures of 18.5--48.0\,kK, mass-loss rates of 10$^{-10}$ to 10$^{-7}$\,M$_{\odot}$\,yr$^{-1}$, and wind velocities between 400 and 2000\,\kms. Only two stars, the two hottest cases, displayed wind velocities larger than 600\,\kms\ and both are amongst the few known X-ray emitters (see previous paragraph). In this paper, masses or radii were not directly determined but they can be reconstructed using the fitted $\log(g)$, $T_{eff}$, and luminosities. The wind velocities appear to be between 1.5 and 3.5 times the escape velocities. \citet{kri19} analyzed in detail two other subdwarfs, one of them showing no trace of wind. The other one (in the binary HD\,49798) has an effective temperature of 45.9\,kK, a mass loss-rate of 2--3$\times10^{-9}$\,M$_{\odot}$\,yr$^{-1}$, and a wind velocity of $\sim$1600\,\kms\ (which is 3.5 times the escape velocity). Again, it is one of the known X-ray emitting subdwarfs. In addition, \citet{gro08} reported on the analysis of the stripped star, of qWR type, in the binary HD\,45166\footnote{\citet{got18} found a continuous sequence, explained by mass, between subdwarf spectra with absorption lines (appearing as sdOs) and stripped star spectra with emission lines (appearing as qWR): such stars are therefore relevant to the problem examined here.}. This qWR star has a temperature of 50\,kK, a strong mass-loss rate (2--3$\times10^{-7}$\,M$_{\odot}$\,yr$^{-1}$), but a low wind velocity (425\,\kms, which is one third of the escape velocity). Data suggest a latitude-dependent wind, but even the fast polar wind would reach only 1300\,\kms. Finally, Drout, G\"otberg, et al. (submitted) have performed a search for stripped-star companions to B/Be stars in the Magellanic Clouds. They found that none of these companions exhibit emission lines, suggesting that all of them actually display low mass-loss rates, in agreement with the above results found in the Galaxy.

Finally, orbits derived for the Be+sdO systems yield masses of 6--13\,M$_{\odot}$ for the Be stars and 0.6--2.\,M$_{\odot}$ for their companions (Table \ref{opt}). We note that this agrees well with the masses derived for \gc\ systems and other Be binaries \citep[see][for a summary]{naz21}, but we will come back to that below. In addition, temperatures of the stripped stars in Be+sdO systems were found to be 34--53\,kK (Table \ref{opt}), similar to those of other studied subdwarfs (see above). In parallel, luminosities of the Be stars in Be+sdO systems (Table \ref{opt}) as well as those of \gc\ analogs \citep{naz18} are far from extreme, being $<10^5$\,L$_{\odot}$.  

In summary, both observations and models suggest that stripped stars with Be companions have low masses, moderate temperatures, and not extremely fast winds. In parallel, the Be stars known to be in binaries with stripped stars or to be \gc\ analogs are found to have around ten solar masses. All this contrasts with the hypotheses of Langer et al. models numbered 4 to 6, which have high masses for both the stripped star and its OB companion (2.6--5.3 and 25.3--35.9\,M$_{\odot}$, respectively), very high luminosities for both of stars ($>10^3$ and $10^5$\,L$_{\odot}$, respectively), very high temperatures for the stripped stars (74--98\,kK), and extremely fast winds ($>3600$\,\kms, assuming $v_{\infty}/v_{esc}=3$). \citet{lan20} acknowledged in their Sect. 5.2 that model 6 in fact represents a WR+O system rather than a Be with a stripped star, but they also considered the models with the most massive and luminous companions as the best candidates for \gc-like emission. Eliminating these unrealistic models strictly limits the available wind luminosities to $L_w=0.5 \dot M v_{\infty}^2 <10L_{\odot}$ (see more in next section). Their models 1--3 agree better with the evolutionary tracks of \citet{got18} and the observed properties of stripped stars in known Be+sdO systems. Only these models will therefore be considered as relevant to the examined problem.

\subsection{The properties of colliding winds}

Colliding winds in massive binaries have now been studied for several decades and their properties over a large range of wavelengths are well known. In the X-ray range, several specific features have been identified \citep[for a review, see][and references therein]{rau16}.

First, it must be recalled that such X-ray bright emission is the exception rather than the rule for massive binaries. For example, in the Carina region, the X-ray emission of 60 O-type stars was studied \citep{naz11}. Only three objects were found to be significantly overluminous: one magnetic star with confined winds and two binaries. Furthermore, the wind luminosity $L_w$ cannot be considered as a good proxy for the X-ray luminosity. Depending on the nature of the collision, the difference between these luminosities can be large. For example, \citet{zhe12} found $L_{\rm X}/L_w\sim10^{-4}-10^{-5}$ for a sample of WR+OB systems. Considering the available $L_w$ for stripped systems ($<10L_{\odot}$, see previous subsection), this renders it difficult to achieve high enough X-ray luminosities (for \gc\ analogs, $\log(L_{\rm X})>31.6$).

Second, the emission arises from the hot plasma generated by the collision: it is thermal in nature. More precisely, it corresponds to optically-thin thermal plasma (e.g. {\sc xspec} models ``mekal'', ``apec''). High-resolution X-ray spectroscopy has confirmed this, by revealing the emission to consist of lines with a faint (bremsstrahlung) continuum \citep[e.g.][]{sch04}. The emission is thus not a blackbody emission, as assumed by \citet{lan20}.

Furthermore, since strong shocks are involved, the temperature of the hot plasma should be evaluated with the well-known Rankine-Hugoniot relationship $kT=\frac{3}{16}mv_w^2$ (and not by $kT=0.5mv_w^2$, as assumed by \citealt{lan20}). In this formula, the wind velocity $v_w$ is the pre-shock velocity perpendicular to the shock: it is not necessarily the terminal wind velocity $v_{\infty}$. Indeed, the stars must be separated enough for their winds to reach their maximum velocities before colliding. In addition, the UV emission from the companion may slow down the wind of a star as it is line-driven by nature (this effect is known as radiative braking/inhibition, see e.g. \citealt{ste94}). Finally, the winds collide in a face-on manner only at the apex of the collision cone, on the line joining the stars' centers. Winds collide more obliquely further away from the line-of-centers, lowering the post-shock temperature. A more representative temperature estimate actually is half the Rankine-Hugoniot value at apex, or $kT=0.6v_w^2$, where $kT$ is provided in keV and $v_w$ is expressed in units of 1000\,km\,s$^{-1}$ \citep{par13}. This leads to a factor of five difference between expected temperatures and those calculated in \citet{lan20}, not even taking the radiative effects into account\footnote{Actually, one could argue that the real difference is larger as \citet{lan20} uses the proton mass $m_p$ in their $kT$ formula while the Rankine-Hugoniot formula applies to the mean particle mass $m$, which is about half the proton mass in an ionized medium.}. To reach temperatures over 5\,keV (and even up to 25\,keV), as observed in \gc\ analogs, shock velocities larger than 3000\,\kms\ (up to 6500\,\kms) are in fact needed, which are not typical wind velocities of stripped stars (see previous subsection). 

Such high temperatures are not typical of colliding wind binaries either, despite the well-known fast winds of massive stars. In observed systems, the X-ray emission is generally harder than the usual intrinsic emission of single non-magnetic massive stars but the hot plasma remains at a much lower temperature than found in \gc\ stars \citep[and references therein]{rau16}. To advocate for extreme temperatures, \citet{lan20} referred to the review of \citet{gag12}. In this review, the vast majority of colliding-wind systems display $kT<5$\,keV (typically 0.5--2\,keV). Only three systems are mentioned with a hotter plasma: CEN\,1A, HD\,5980, and Brey\,16. The first system was studied in detail using \xmm\ by \citet{mer13}: the hottest plasma component had a temperature of 3.8\,keV. The difference with the previous \ch\ value most probably comes from pile-up effects, known to artificially shift X-rays towards higher energies. The second system, analyzed by \citet{naz02}, is embedded in patchy soft X-rays from a supernova remnant. A clean spectrum of the system is difficult to extract, even with \ch, and possibly explains the anomalous temperature. The X-ray properties from the last system were reported by \citet{gue08}. Only few counts were available for this system and the median energy of the recorded X-rays is about 2\,keV. A tentative\footnote{From \citet{gue08}: {\it ''We note that LMC-WR 19, 20, and 99 have small numbers of counts detected, and consequently the quality of their spectral fits is poor.''} - LMC-WR19 is Brey\,16.} fit does formally land on a plasma temperature of 7\,keV, but with a wide confidence interval (from 2 to 22\,keV at 1$\sigma$). Clearly the faintness of the source did not allow reliable results to be reached. Furthermore, in all three cases, the well-known trade-off between absorbing column and plasma temperature may also skew the results somewhat. Therefore, the basis for a claim of extremely high temperatures associated with colliding winds appears unsubstantiated.

Besides, \citet{lan20} themselves also advocated in their Sect. 5.2 that temperatures of their model need to be lowered to $\sim$1.3\,keV to be reconciled with the observed hardness ratios of \gc\ analogs. This is clearly at odds with the observed properties of \gc\ X-ray spectra (which show $kT>5$\,keV), but may come from a combination of the inadequate choice of a blackbody model with an inadequate definition of hardness ratios\footnote{These authors compare observed ratios which are ratios of fluxes integrated in two wide energy bands (0.5--2.0\,keV and 2.0--10.\,keV) with ratios of blackbody fluxes at two specific energies (1 and 5\,keV).}.

A last defining characteristic of colliding winds is their variability. This variability is locked to the orbital phase, for a large range of periods (from a few days to several years) - repeatability was verified for several systems \citep[e.g. in 20 cycles of WR\,21a, see][]{gos16}. It was linked to several origins: change in shock strength due to the varying separation in eccentric binaries \citep[e.g.][]{naz12}, change in absorption along the line-of-sight as the binary revolves \citep[e.g.][]{wil95}, or eclipse of the colliding wind region \citep{lom15}. Such variations are now rather well understood and can be reproduced by models \citep[e.g.][]{pit10}. Such features appear in stark contrast with the X-ray properties of \gc\ analogs. These objects show `flare'-like variations on short timescales (seconds, see \citealt{smi98}) and smoother long-term variations, but not linked with the known orbital periods (see \citealt{mot15} for \gc, and \citealt{naz19piaqr} for $\pi$\,Aqr).

Recently, \citet{che21} have asserted to bring support to the Langer et al. scenario by examining the consequences of the detection of a wind-blown bubble around \gc. Amongst other things (see also below), they stated that wind-wind collisions can produce hot plasma but, as seen before, actual colliding winds never reach the typical plasma temperatures of \gc\ analogs. The authors also added that short-term `flare'-like variability of \gc\ could come from the intrinsic variability of winds (i.e., not linked to colliding winds). While some variations of that embedded wind shock emission are indeed expected \citep{osk01}, they should occur on timescales related to the wind expansion, not on the very short timescales observed for the X-ray lightcurve of \gc\ (down to 4\,s, \citealt{smi98}). Moreover, even the best observational datasets of single massive stars did not reveal such short-term X-ray variability \citep{naz13}. Because of the high number of wind clumps, the natural embedded wind shock variability is actually smoothed out when observing the whole wind output, leading to a constant level for the observed emission. Besides, by nature, those intrinsic X-rays are soft, not hard. \citet{che21} finally explain the long-term X-ray variations as usual phase-dependent changes seen in colliding winds, linking them to a varying stellar separation or absorption. Unfortunately, \gc\ does not display any flux variations in phase with its orbital period in the X-ray range \citep{rau22}, its orbit is circular \citep[e.g.][]{nem12}, and high-absorption events are not only unrelated to orbital phase but are sharp events occurring on rather short timescales \citep{smi12,ham16,rau22} whereas absorption effects in colliding winds are smooth and phase-dependent \citep[and references therein]{rau16}. 

\subsection{Additional features}

Regarding the incidence of \gc\ systems, \citet{lan20} conclude in their Sect. 5.1.3 that \gc\ stars should represent at most 10\% of X-ray binaries with a Be component, i.e. there should be many more Be X-ray binaries than \gc\ stars. While this statement may be true if one takes a quick look at current X-ray binaries catalog numbers, it must be remembered that X-ray binaries are detected from afar while most \gc\ stars were found by chance or from limited surveys of nearby stars. In fact, few Be stars (other than those in X-ray binaries) have been observed in the X-ray range \citep[e.g.][]{naz18}. Therefore, the exact incidence of the \gc\ phenomenon is unknown. However, even if distance-limited samples remain to be investigated, the number of current \gc\ detections already indicates that the incidence of the \gc\ phenomenon goes well over, not well below, that of Be X-ray binaries in the Be population at large (see a thorough discussion in Sect. 6.1 of \citealt{smi17}).

\citet{lan20} also discuss the impact of structured winds on the X-ray emission of massive stars. Indeed, discrete emission components (DACs) in spectral lines have been spotted in several \gc\ analogs, notably in \gc\ itself \citep{cra00}, as they have been in many OB stars. These DACs are produced in large-scale wind structures thought to ultimately arise from surface features \citep{cra96}. Such ``spots'' were indeed found in the analysis of the high-quality, high-cadence photometry available for some massive stars presenting DACs \citep{ram14,ram18}. X-rays associated with large-scale wind structures have also been studied for a sample of massive stars \citep{osk01zetoph,rau15,mas19,nic21}. They produce variability of a limited amplitude (10--20\%), with a specific timescale (set by rotation), and only at energies where intrinsic X-rays arise, i.e. in the soft band. Those properties disagree with those observed for \gc\ stars. Furthermore, \citet{lan20} envisaged that a collision between the Be wind and the stripped star wind could be modulated on a ``broad range'' of timescales because of such features in the Be star wind. As acknowledged by \citet{lan20}, DACs are ubiquitous amongst massive stars. Therefore, such variations should already have been detected in colliding winds systems, but this is not the case. This leaves little support for DACs as the origin of the `flare'-like behaviour of \gc\ analogs.

\begin{figure*}
  \begin{center}
    \includegraphics[width=5.5cm]{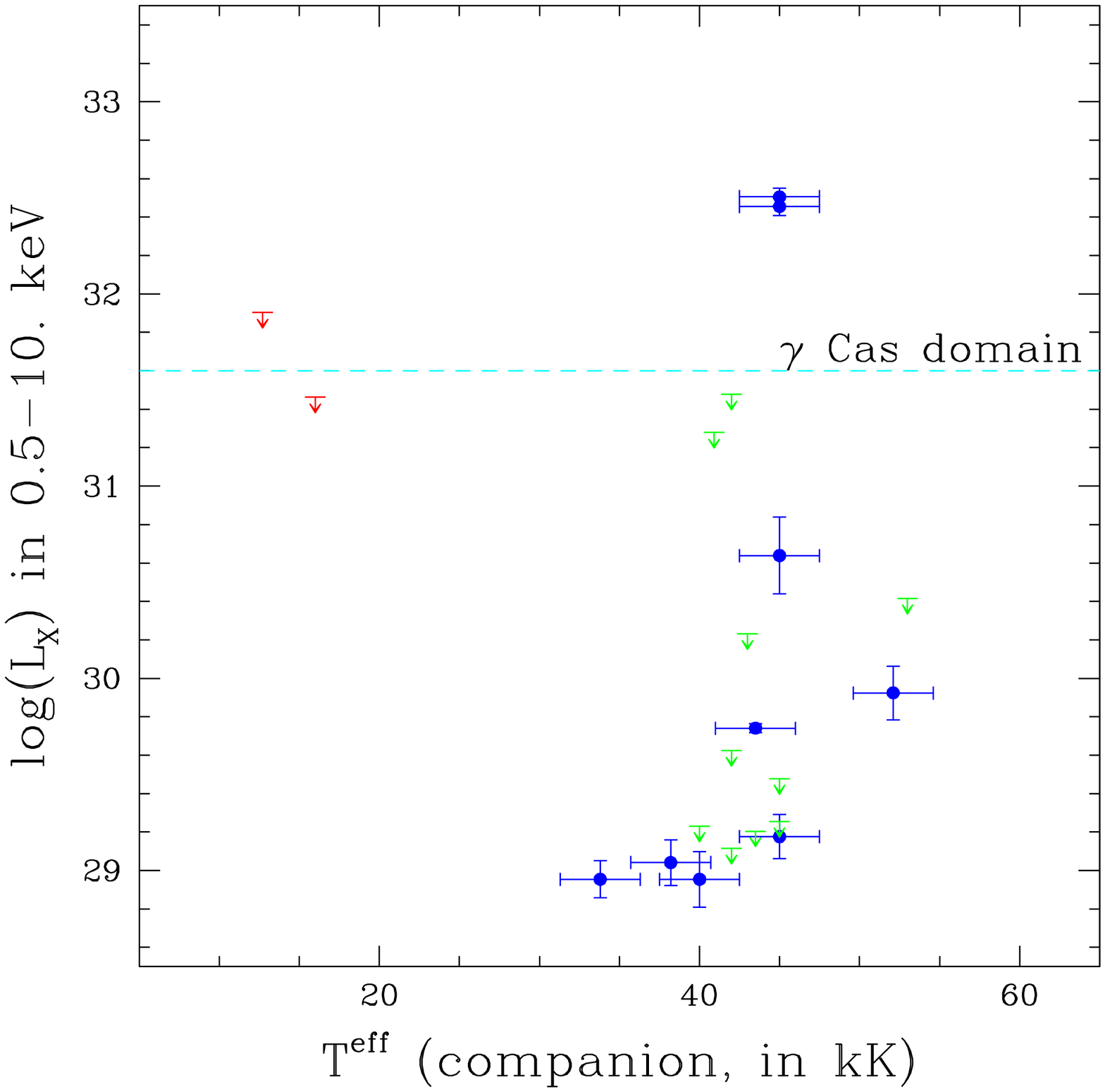}
    \includegraphics[width=5.5cm]{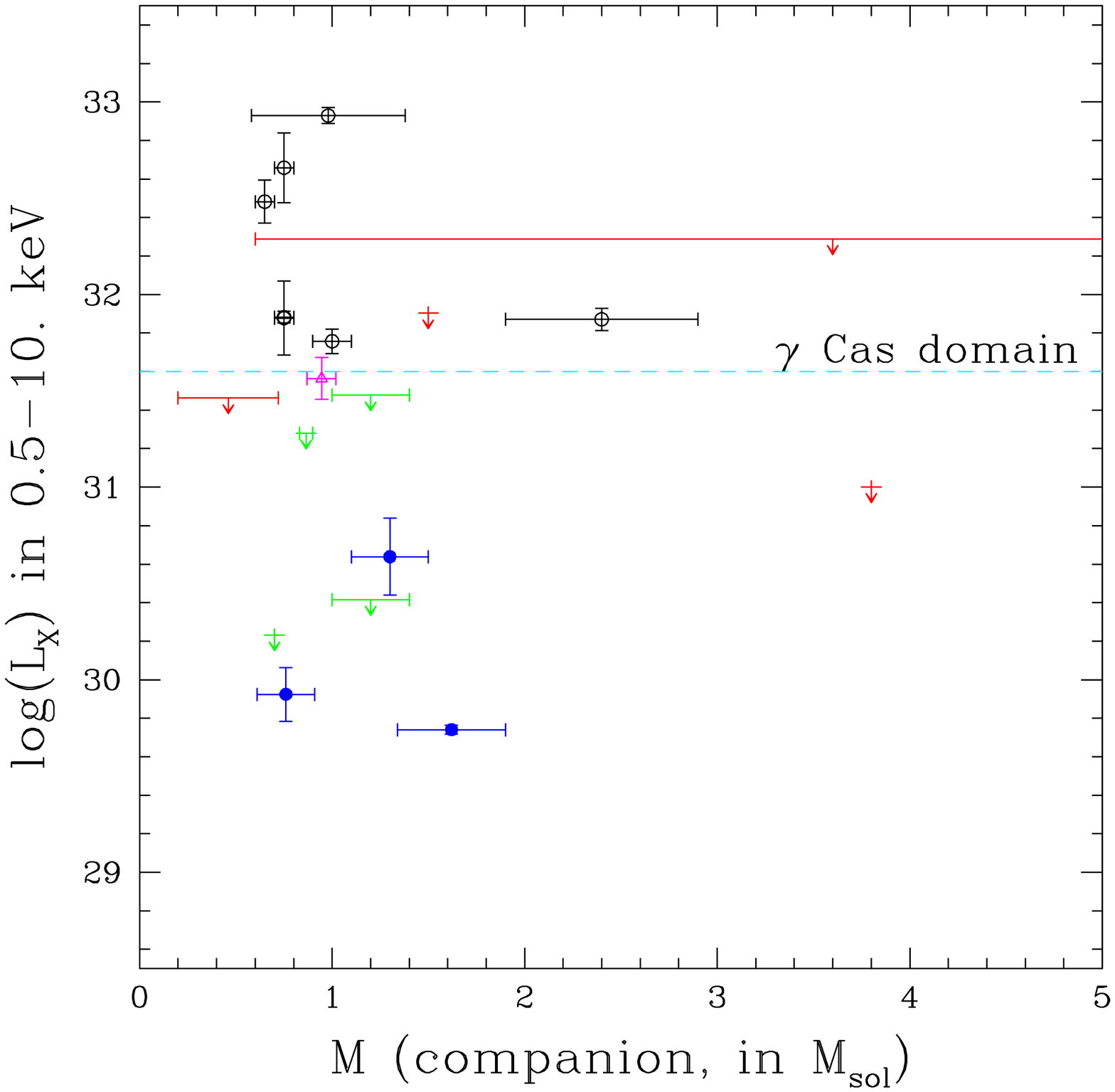}
    \includegraphics[width=5.5cm]{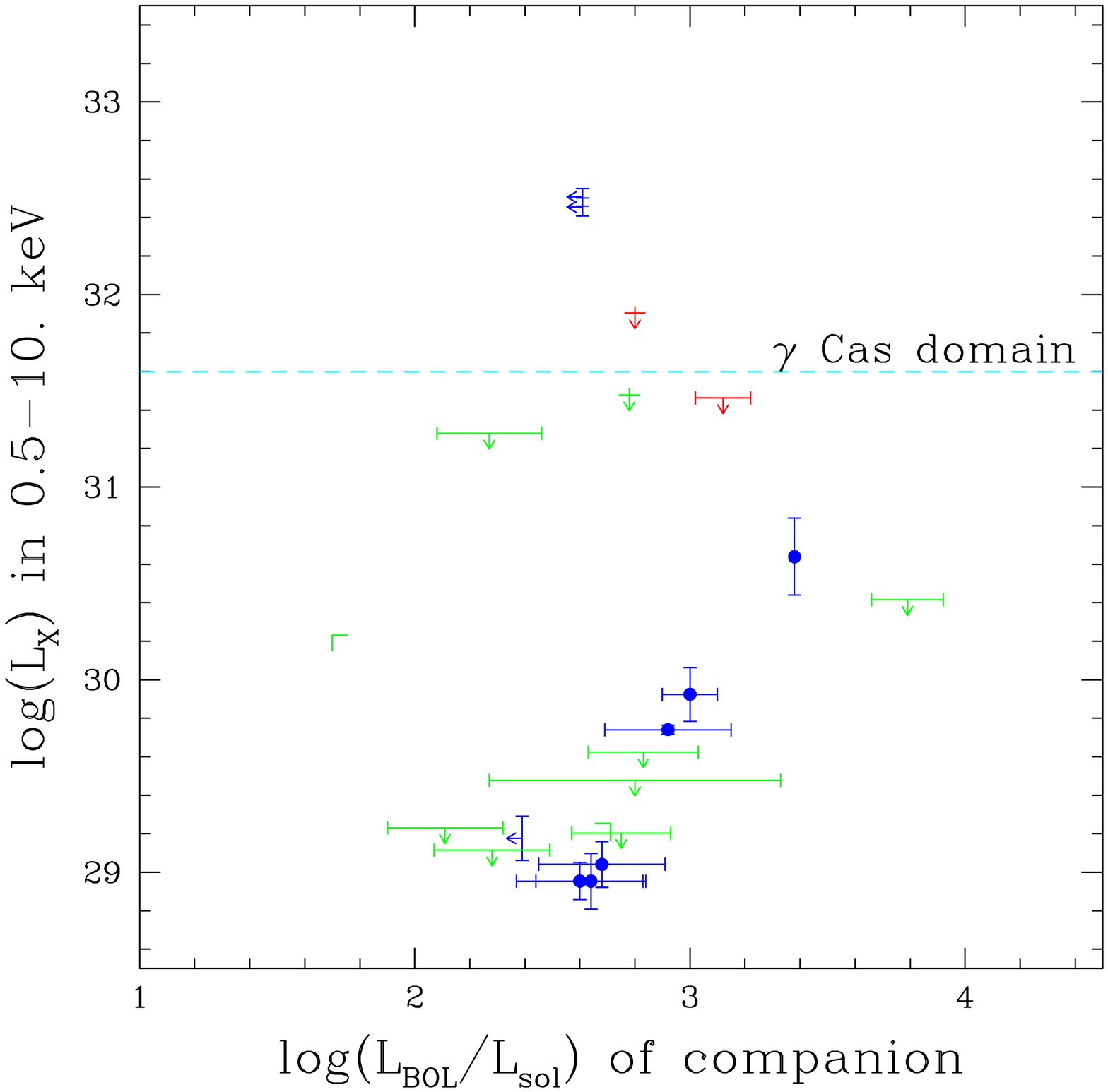}
  \end{center}
  \caption{X-ray luminosities (or upper limits on X-ray luminosities) as a function of temperature, mass, and bolometric luminosity of the companion. Blue circles are used for Be+sdO systems with detected X-ray emissions, green symbols for undetected Be+sdO systems, a magenta triangle for the Be binary with detected X-rays, red symbols for the other undetected Be binaries, and black symbols for \gc\ stars reported in \citet{naz18,naz20,naz21}.}
\label{param}
\end{figure*}

\begin{figure}
  \begin{center}
    \includegraphics[width=7cm]{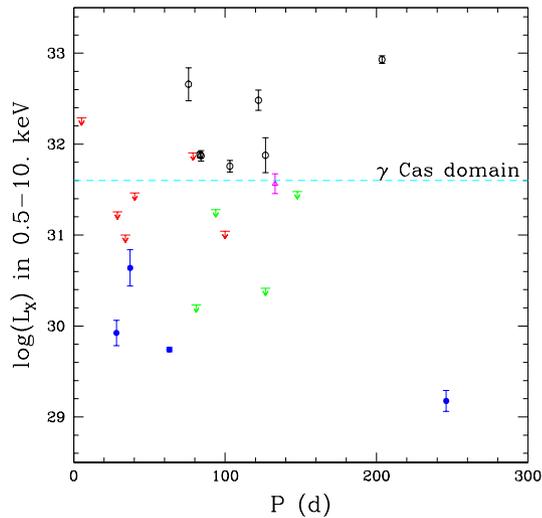}
  \end{center}
  \caption{As in Fig. \ref{param} but with respect to the orbital period of the Be systems. }
\label{param2}
\end{figure}

If a disk-wind collision occurs, one could expect some deviation from pure symmetry in the disk of the Be star \citep{lan20}. Asymmetries (e.g. one-sided arms) are often reported for Be stars from analyses of the H$\alpha$ line profiles. Indeed, the relative amplitude of the emission peaks changes with time (the so-called $V/R$ modulation) and this can be related to structures in the disks. Such disk features are thought to be triggered by the Be star oblateness \citep{pap92} or by companions \citep{pan18}.  \citet{lan20} quotes the case of $\pi$\,Aqr as an example of asymmetric disk bringing (indirect) support to their scenario. However, while its disk strongly developed in recent years, the $V/R$ modulation observed in $\pi$\,Aqr disappeared but not its \gc\ character \citep{naz19piaqr,naz19b}. It may be important to note in this context that the disk estimated size remains well below the binary separation, so that an engulfing of the companion (as proposed in some cases by \citealt{lan20}) cannot apply here. Furthermore, \citet{lan20} also mentioned the possibility for the collision zone to produce H$\alpha$ emission. If that occurs, it means that the collision is efficiently cooled and has become highly radiative. In such a case, the plasma temperature and its associated X-ray emission would drastically drop, which becomes a problem for the envisaged scenario (see also above discussion on $L_{\rm X}/L_w$). In addition, in both $\pi$\,Aqr \citep{naz19piaqr} and \gc\ itself \citep{rau22}, tomographic mapping of the H$\alpha$ emission did not reveal any stable structure linked to a companion, as would be expected in such a case.

\citet{che21} found a cavity of radius 2.1--3\,pc around \gc\ expanding with a velocity of 5\,\kms\ and a kinematic timescale of 0.3\,Myr. Using wind properties derived for the Be star in the literature, they estimated the wind luminosity $L_w$ and found it to be in good agreement with that derived from wind-blown bubble properties. This implies that the Be star {\it alone} can carve this bubble. Then the authors used two $L_{\rm X}-L_w$ relations established for embedded wind shocks (i.e. for massive stars without X-ray bright colliding winds). They concluded that the derived X-ray luminosity can be reconciled with the observed one of \gc, despite the facts that the intrinsic X-rays from massive stars are soft and follow a specific $L_{\rm X}-L_{\rm BOL}$ relation, in sharp contrast to the known X-ray properties of \gc. Note that, in all this, nothing relates to the Langer et al. scenario (i.e. presence of a stripped companion with strong mass-loss, presence of a disk-wind collision) therefore this scenario cannot be supported by the bubble discovery. Finally, {\it WISE} data are presented as evidence for binarity, quoting previous studies. However, these studies do not tell the same story: \citet{bod18} found those data to indicate a morphological classification ``not classified'' for \gc, while \citet{lan20} found these data providing ``weak evidence'' for the presence of a bow shock (which could have been linked to a kick after a supernova explosion - this is actually one argument used by Langer et al. to reject the neutron star nature of the companions in \gc\ and $\pi$\,Aqr). Neither paper considers {\it WISE} data as support for the binarity of \gc.

\subsection{Testing the Langer et al. model}

While the analogy of \gc\ stars with known colliding-wind systems remains elusive, it is nevertheless of great interest to examine how X-ray properties of Be stars depend (or not) on binarity. Amongst the cases studied in the X-ray range, there are eight \gc\ stars which are known binaries \citep{naz21}, in addition to the 18 Be+sdO systems and 7 other Be binaries studied in this work. Figures \ref{param} and \ref{param2} compare the X-ray luminosities of these systems to the temperature, mass, and luminosity of the companions and to the orbital period of the systems (see Table \ref{opt} or \citealt{naz21} for their values).

\citet{lan20} expected a higher X-ray luminosity for systems with more massive companions, with only the most massive companions leading to a \gc\ appearance: ``only \gc\ binaries with sufficiently massive helium-star companions are predicted to have detectable X-ray fluxes'' (their Sect. 5.1), ``Other predictions resulting from the given ansatz are that \gc\ stars might have rather massive helium-star companions'' (their Sect. 5.3). This can now be tested with observations. The middle panel of Fig. \ref{param} graphically shows the absence of correlation between the companion mass and the X-ray luminosity in our sample. It also fails to demonstrate a clear mass segregation in relation to the presence or absence of the \gc\ character. To this, it must be added that the \gc\ binaries studied before have not revealed companions with particularly high masses. For example, the well studied case of \gc\ led to a mass estimate around 1\,M$_{\odot}$ \citep{nem12,smi12}. In fact, there does not seem to be a significant difference between companions in \gc\ systems and companions in other Be binaries \citep[and Fig. \ref{param}]{naz21}. The Langer et al. prediction thus appears falsified.

One could also expect a stronger disk-wind collision if the companion's wind is stronger. Since the wind depends on temperature or, equivalently, on the bolometric luminosity \citep{kri16,vin17}, the left and right panels of Fig. \ref{param} also compare the X-ray luminosities to these parameters, when known. Some marginal correlation may be seen by eye for the five X-ray detections without a \gc\ character, but (1) this relies on so few systems that spuriosity cannot be excluded and (2) this neglects both the upper limits and the \gc\ case. In particular, it is important to note that the sole \gc\ analog amongst the presented sample is HD\,157832 (the sole blue symbol above the \gc\ border in Fig. \ref{param}). This target harbours a faint and not-very-hot stripped star companion \citep{wan21}, notably contradicting Sect. 5.2 of \citet[``the known \gc\ stars should correspond to the most luminous models that predict the \gc\ phenomenon.'']{lan20}. Combining its luminosity limit (Table \ref{opt}, \citealt{wan21}) with formulae of \citet{kri16} or \citet{vin17} yields $\log (\dot M)<-9.99$ or --9.75, respectively, for this stripped star. This implies that the \gc\ character may well exist with a stripped-star companion having a tenuous stellar wind, at odds with a disk-collision scenario. 

One could also consider the separation between the binary components as a potentially important parameter. Indeed, with wider separations, the disk must extend farther before the collision with the companion's wind takes place. The disk material would then be more tenuous, and the collision would then be weaker. Figure \ref{param2} therefore compares the X-ray luminosities to the orbital periods. No obvious correlation stands out. For example, the two long-period systems (\gc\ and HD\,191610) harbour X-ray luminosities at both extremes of the range. There is one caveat, however: Be disks are known to vary and not all disks may be large enough at the time of the X-ray observations to reach the companions. In this context, Table \ref{donx} also lists the equivalent width of the H$\alpha$ line close to the time of the X-ray observations, when available. Most of the strongest disk emissions are found for systems which are not \gc\ in character! Furthermore, it may be remembered that the Be disks are often truncated \citep{kle19}. For Be X-ray binaries \citet{oka01} showed that this truncation often occurred at about one third of the orbital separation. All this implies that little disk material would be present in the companion's neighbourhood, even when the disk appears massive. Moreover, dedicated monitorings have shown that the \gc\ character may exist even with a very limited disk (see the cases of $\pi$\,Aqr, HD\,119682, and V767\,Cen in \citealt{naz19piaqr,naz22}). All these pieces of evidence provide additional arguments against a disk-collision scenario. 

Finally, since orbital solutions are known in some cases (Table \ref{opt}), one may wonder at which phases the dedicated X-ray observations were taken. ALS\,8775, HD\,41335 (second observation), and HD\,58978 were observed as the companion was in front of the Be star (i.e. at superior conjunction of the Be star). HD\,10516, HD\,41335 (first observation), and HD\,200120 were observed near quadratures. HD\,37202 was observed just after superior conjunction of the Be star. HD\,55606 was observed between quadrature and inferior conjunction of the Be star. HD\,194335 was observed when the Be star was in front. In all cases but the latter one, there is no reason to expect an X-ray source close to the companion to be eclipsed by the Be star or its disk. Furthermore, even in the latter case, such an eclipse could be avoided since its inclination is ``intermediate'' \citep{kle21}. One can thus conclude that there was no specific reason for a \gc\ character, if present, to remain hidden in those systems. 

\section{Conclusion}
Using \xmm, \ch, and \sw\ observations, we have examined the X-ray emission of a set of Be binaries: 18 known to harbour a stripped-star companion and 7 for which the companion nature is debated. The derived X-ray properties appear fully in line with those found for other samples of Be stars, revealing no specific effect of the presence of these companions on the X-ray emission. In particular, 21 systems display faint (and soft) X-rays while two others clearly appear as \gc\ analogs (one of them, HD\,37202, being a new detection). This incidence rate appears similar to that seen for other Be samples. No relationship could be found between the X-ray luminosities and the periods of the systems or the properties of the companions. In particular, there is a single case of a \gc\ star with constrained companion's luminosity, HD\,157832, and this companion appears faint. The \gc\ character may thus exist even if the companion ejects little wind. In addition, the link between the \gc\ character and the companion's mass predicted by \citet{lan20} is not verified. Combined with problematic analogies with wind-wind phenomenology and stripped-star properties, this makes a disk-wind collision an unlikely explanation for the \gc\ phenomenon. 

\section*{Acknowledgements}
The authors thank the referee for his comments that helped improved the paper. Y.N. and G.R. acknowledge support from the Fonds National de la Recherche Scientifique (Belgium), the European Space Agency (ESA) and the Belgian Federal Science Policy Office (BELSPO) in the framework of the PRODEX Programme (contracts linked to XMM-Newton). M.A.S. acknowledges support from \ch\ grant \#362675. ADS and CDS were used for preparing this document. This work has made use of the BeSS database, operated at LESIA, Observatoire de Meudon, France (http://basebe.obspm.fr).

\section*{Data availability}
The BeSS, \xmm, \sw, and \ch\ data used in this article are available in their respective public archives.

\appendix

\section{The optical emission of HD\,37202 ($\zeta$\,Tau)}

\begin{figure*}
  \begin{center}
    \includegraphics[width=7.cm]{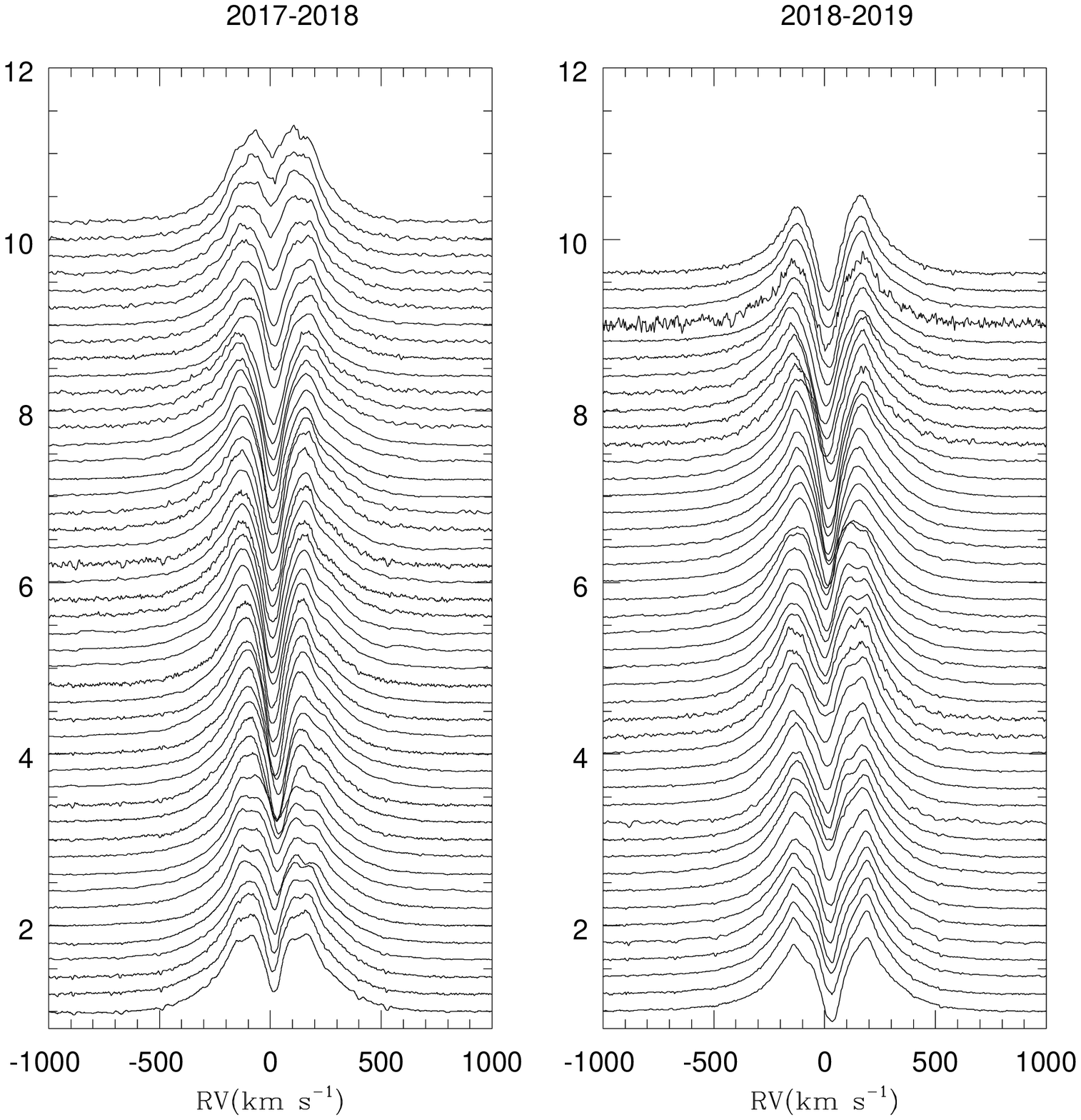}
    \includegraphics[width=7.cm]{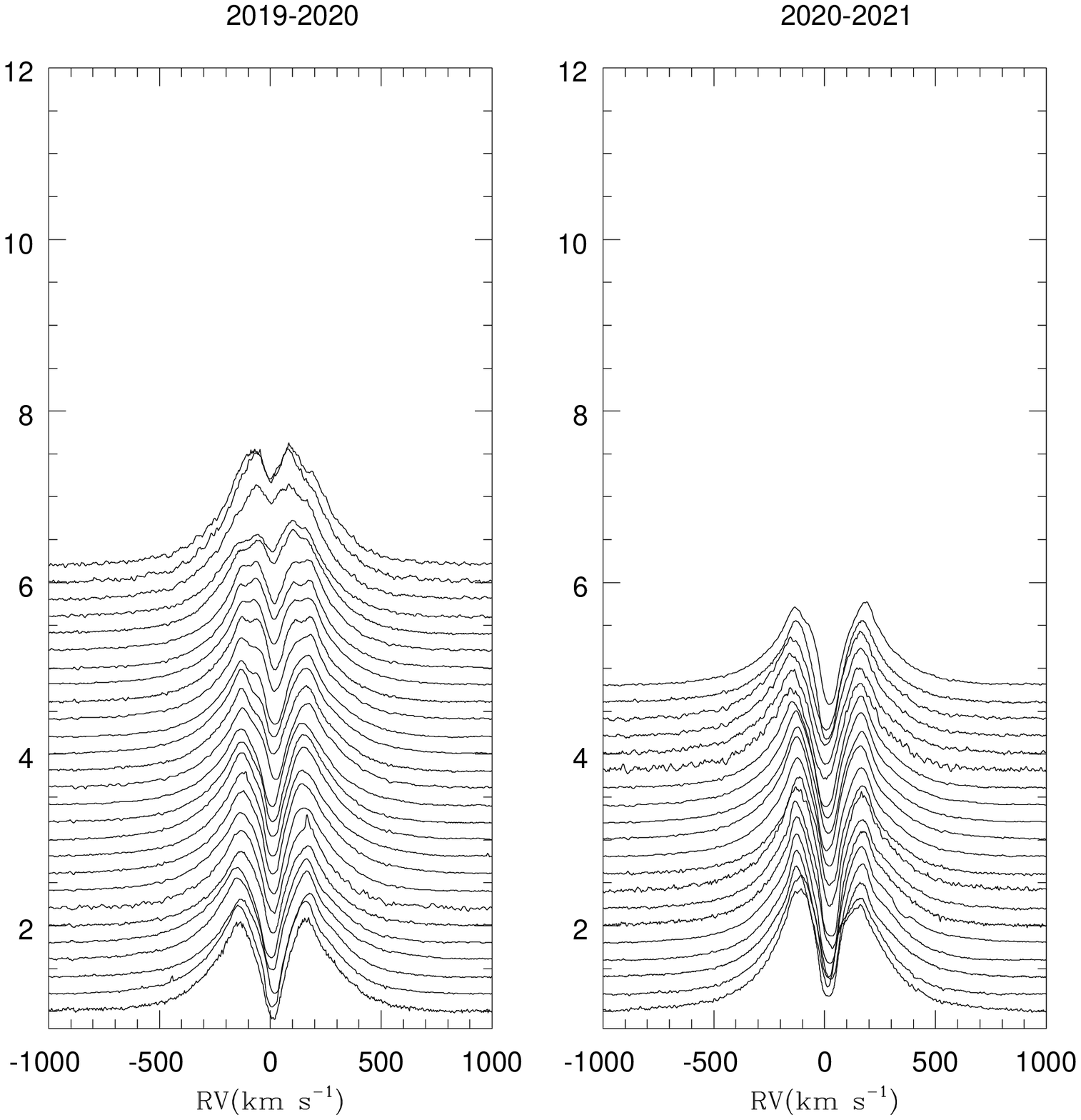}
    \includegraphics[width=3.5cm, bb=20 145 305 720, clip]{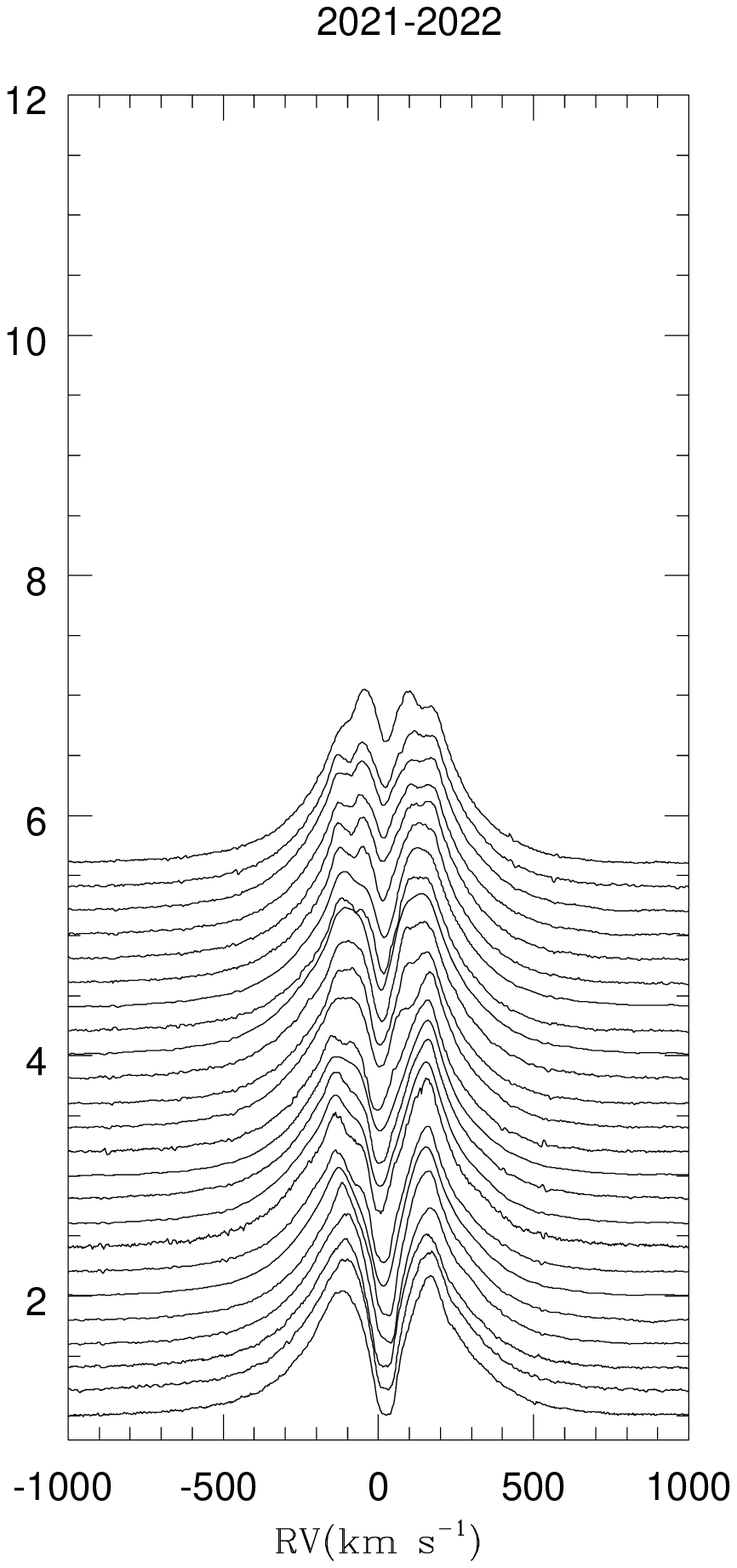}
  \end{center}
  \caption{Evolution of the profile of the H$\alpha$ line in the spectrum of HD\,37202 over the latest five observing seasons. Time is running downwards.}
\label{ztprof}
\end{figure*}

\begin{figure}
  \begin{center}
    \includegraphics[width=8.5cm]{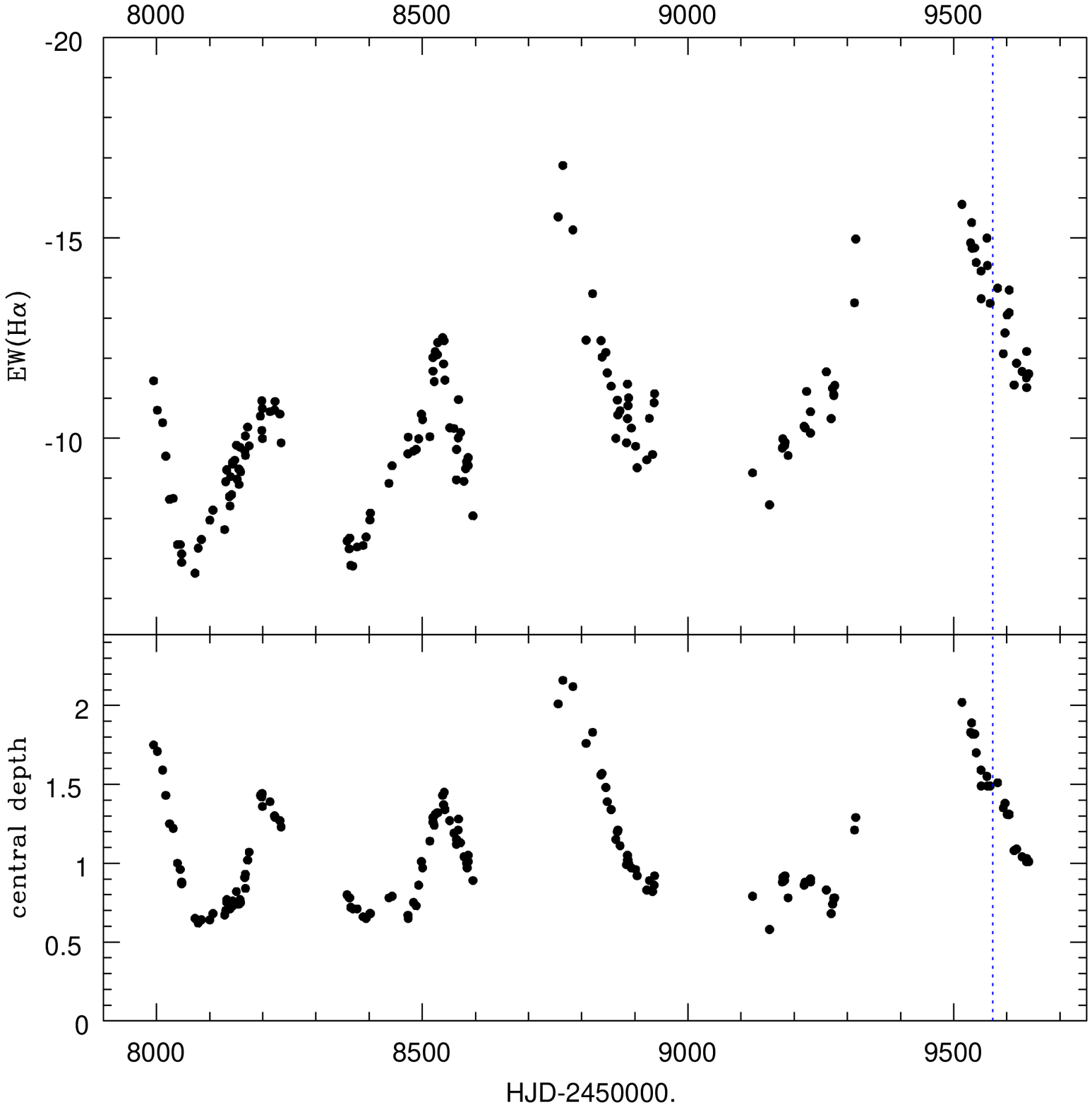}
  \end{center}
  \caption{Evolution with time of the $EWs$ of the H$\alpha$ line (evaluated between --600\,km\,s$^{-1}$ and 600\,km\,s$^{-1}$) and of the normalized amplitude of the lowest point in the central absorption (a value of unity implying a lowest point appearing at the continuum amplitude, values $<1$ implying a central absorption reaching below the continuum level, and values $>1$ absorptions remaining above the continuum level). The time of the \ch\ observation of HD\,37202 is marked by a dashed blue line.}
\label{ztew}
\end{figure}

\begin{figure}
  \begin{center}
    \includegraphics[width=8.5cm]{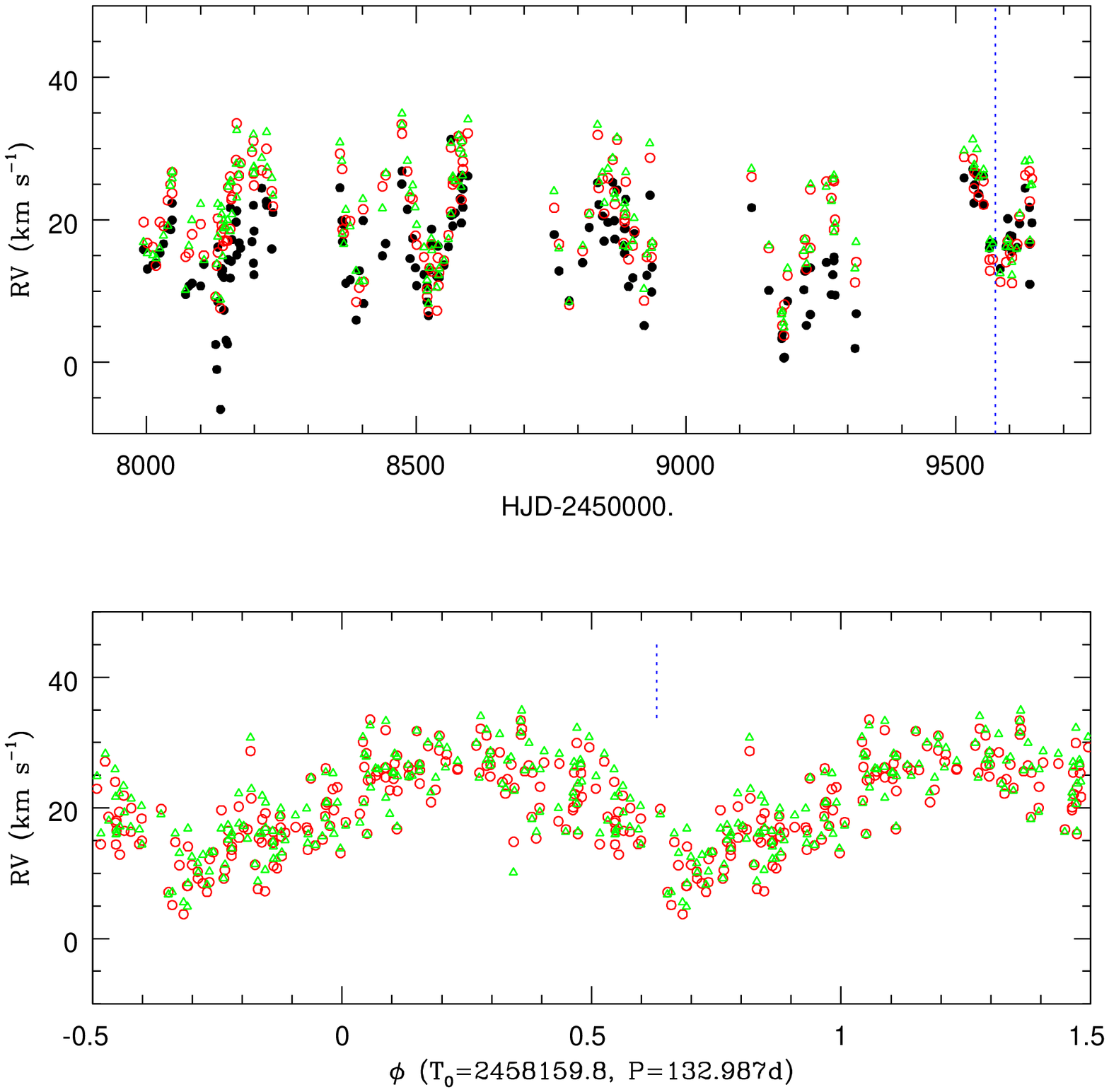}
  \end{center}
  \caption{Evolution with time and phase of the $RVs$ of the H$\alpha$ line derived for HD\,37202. Black dots correspond to 1st-order moments, red circles to velocities evaluated using the mirror method, and green triangles to velocities from the double-Gaussian method. The time and phase of the \ch\ observation is marked by a dashed blue line.}
\label{ztrv}
\end{figure}

Thanks to its brightness, HD\,37202 ($\zeta$\,Tau) has been observed for a very long time. Its binarity is known since \citet{ada05} and the high inclination of the disk of its Be component was long assumed because of the presence of shell-type profile (and demonstrated through interferometric measurements since at least \citealt{qui97}). In parallel, long-term spectral variations can be spotted for this star due to a changing disk, as in other Be stars. In this context, a unique decadal monitoring covering 1997--2008 revealed a cyclic behaviour of duration $\sim$1400\,d \citep[e.g.][]{ste09}. Focusing on H$\alpha$, the line profile first displayed a very asymmetric shape, with the violet peak largely dominating. Then the amplitude difference between the peaks decreased and the profile, because of the central absorption, took a clear ``shell'' aspect. In a third phase, the profile became again asymmetric, but this time with the red peak dominating. Finally, the line returned to a symmetrical situation but not with a shell appearance - instead, a pure emission line with multiple small subpeaks could be seen.

This behaviour was interpreted in several ways. \citet{car09} performed 3D NLTE radiative transfer modelling. They could reproduce the observed profile variations by considering a viscous decretion disk with a one-armed spiral. \citet{sch10} rather interpreted the profile variations by considering a precessing disk.

Over that 1997--2008 decade, the average equivalent width of the line remained quite stable near --20\AA, but the emission strength subsequently dropped to reach a minimum in early 2013 \citep[see Fig. 2 in][]{pol17}. The disk has since (partially) recovered, but the behaviour seems to have changed. Typically, the target is observed from Fall to Spring and the previous study \citep{pol17} stopped in Spring 2017. We therefore decided to focus on data taken since then, i.e. in the last five observing seasons.

Spectra were downloaded from the BeSS database \citep[and http://basebe.obspm.fr]{nei11}: five low-resolution ($R<6500$) spectra were discarded as most spectra have $R=10\,000-20\,000$. Signal-to-noise ratios ranged from 20 to 700, with a typical value of 150--350. In total, 162 spectra were examined, splitted in the following manner: 47 in 2017--8, 44 in 2018--9, 27 in 2019--20, 20 in 2020--21, and 24 in 2021--2. These spectra were corrected for telluric absorption within {\sc IRAF} and then normalized using low-order polynomials through the same wavelength windows. 

Figure \ref{ztprof} displays the observed profile variations of the H$\alpha$ line. The strongly asymmetric profiles, which were common in 1997--2008, have now totally disappeared. Instead, the profiles remain symmetric, with little V/R changes. The only source of variations comes from the changing depth of the central absorption. Figure \ref{ztew} provides the evolution of $EWs$ and of the normalized amplitude of the lowest point in the central absorption (see also Table \ref{tablerv}). Both parameters show a good correlation, showing that the $EW$ variations are indeed mostly driven by the changing absorption depth.

The profile variations appear to repeat, albeit not in an identical way. The $EWs$ were indeed larger in the last three seasons than in the first two. Moreover, the duration of these cycles does not remain constant as there are 200--400\,d between two consecutive minima or two consecutive maxima. Such timescales appear to be in agreement with preliminary findings from \citet{pol17} but they are well below the previously reported cycle length of 1400\,d, however. It is thus difficult to explain the cyclic behaviour of HD\,37202 by precession, since the period of that phenomenon would not change over a few decades. If disk structure are instead responsible for the profile variations, then the observations indicate a change in the disk geometry over time, with a recent disappearance of the spiral arm spotted in 1997--2008. 

Finally, we have evaluated the radial velocity $RV$ of the H$\alpha$ line. To this aim, we have first computed the first-order moment of the line. However, since the core of the line varies quite a lot, this estimate may be biased. Therefore, we also calculate the $RVs$ using two other methods. The mirror method compares the blue wing to the mirrored (i.e. reversed in velocities) red wing, for several shifts. It has been used notably for \gc\ itself \citep{nem12}. The wings were considered between normalized amplitudes 1.16 and 1.5, which enables us to totally avoid the central absorption. The double-Gaussian method, also used on \gc\ \citep{smi12}, correlates the line profile with a function composed of two Gaussians with identical widths (here, 15\,km\,s$^{-1}$), opposed amplitudes and centers (here, centers were set at $\pm300$\,km\,s$^{-1}$). Both methods avoid considering the central core, which is often strongly affected by disk variations in Be stars. The results of these $RV$ determinations are provided in Table \ref{tablerv} and shown in Fig. \ref{ztrv}.

A period search, using notably a modified Fourier algorithm (see e.g. \citealt{naz21}), was applied to the velocities. It yielded a period of 133.3$\pm$1.1\,d for the mirror $RVs$ and 132.5$\pm$1.1\,d for the double-Gaussian $RVs$. These periods agree with previous determinations but are less precise. We therefore adopted the same period value as \citet{rud09}: $P=132.987$\,d. Orbital solutions were then calculated for this period. When considering excentric solutions, the best-fit eccentricity does not appear to be significant: $e=0.09\pm0.07$. Circular solutions are thus favored, with parameters: $K=7.4\pm0.8$\,km\,s$^{-1}$, $v_0=21.2\pm1.1$\,km\,s$^{-1}$, and $T_0=2\,458\,159.8\pm3.0$ (conjunction with the Be primary in front). Those values agree well with the orbital solution of \citet{rud09}.

\begin{table*}
  \scriptsize
  \caption{Characteristics of the H$\alpha$ line observed for HD\,37202 since Fall 2017. \label{tablerv}}
  \begin{tabular}{lccccc lccccc lccccc}
    \hline
$HJD$ & $EW$(H$\alpha$) & depth & \multicolumn{3}{c}{$RV$(km\,s$^{-1}$)} & $HJD$ & $EW$(H$\alpha$) & depth & \multicolumn{3}{c}{$RV$(km\,s$^{-1}$)} & $HJD$ & $EW$(H$\alpha$) & depth & \multicolumn{3}{c}{$RV$(km\,s$^{-1}$)}\\
$-2.45e6$& (\AA) & & M1 & mirror & 2G & $-2.45e6$& (\AA) & & M1 & mirror & 2G & $-2.45e6$& (\AA) & & M1 & mirror & 2G\\
\hline                                               
7994.666 & -11.4 &  1.75 &  15.8 &  19.7 &  16.9 & 8394.557 &  -7.5 &  0.65 &  12.8 &  10.5 &  11.3 & 8887.393 & -10.8 &  1.02 &  18.8 &  20.7 &  20.6 \\ 
8001.662 & -10.7 &  1.71 &  13.1 &  16.7 &  15.3 & 8401.580 &  -8.0 &  0.68 &  19.9 &  21.5 &  22.9 & 8888.414 & -11.0 &  1.00 &  22.9 &  25.3 &  26.7 \\ 
8011.663 & -10.4 &  1.59 &  13.7 &  16.2 &  15.0 & 8402.606 &  -8.1 &  0.68 &   8.2 &  11.3 &  11.4 & 8893.424 & -10.3 &  0.97 &  10.6 &  14.5 &  16.1 \\ 
8017.674 &  -9.6 &  1.43 &  13.9 &  13.6 &  14.6 & 8437.445 &  -8.9 &  0.78 &  14.9 &  24.7 &  21.6 & 8901.446 &  -9.8 &  0.96 &  11.9 &  16.3 &  17.0 \\ 
8024.635 &  -8.5 &  1.25 &  15.3 &  19.7 &  16.0 & 8443.469 &  -9.3 &  0.79 &  16.7 &  26.3 &  26.5 & 8904.327 &  -9.3 &  0.92 &  18.1 &  18.4 &  20.3 \\ 
8031.564 &  -8.5 &  1.22 &  16.6 &  19.1 &  17.7 & 8473.408 &  -9.6 &  0.67 &  25.0 &  33.4 &  33.3 & 8922.378 &  -9.5 &  0.83 &   5.2 &   8.7 &  10.3 \\ 
8039.554 &  -7.3 &  1.00 &  18.8 &  22.7 &  19.1 & 8473.693 & -10.0 &  0.65 &  26.8 &  32.1 &  34.9 & 8927.373 & -10.5 &  0.89 &  12.2 &  14.8 &  15.1 \\ 
8044.637 &  -7.3 &  0.96 &  18.6 &  25.0 &  24.7 & 8483.687 &  -9.7 &  0.75 &  21.5 &  26.8 &  28.3 & 8933.343 &  -9.6 &  0.82 &  23.4 &  28.7 &  30.7 \\ 
8047.493 &  -7.1 &  0.88 &  22.3 &  26.7 &  26.7 & 8488.687 &  -9.7 &  0.73 &  14.6 &  23.1 &  23.8 & 8936.348 & -10.9 &  0.86 &   9.9 &  14.7 &  16.0 \\ 
8047.529 &  -6.9 &  0.87 &  19.9 &  23.7 &  25.8 & 8493.429 & -10.0 &  0.86 &  17.4 &  22.9 &  24.8 & 8937.314 & -11.1 &  0.92 &  13.4 &  16.6 &  16.9 \\ 
8072.540 &  -6.6 &  0.65 &   9.5 &  14.8 &  10.1 & 8498.416 & -10.6 &  1.01 &  13.2 &  17.8 &  21.7 & 9121.577 &  -9.1 &  0.79 &  21.7 &  26.0 &  27.1 \\ 
8078.569 &  -7.3 &  0.62 &  10.7 &  15.3 &  16.4 & 8500.774 & -10.5 &  0.97 &  10.8 &  16.5 &  19.3 & 9153.556 &  -8.3 &  0.58 &  10.1 &  16.0 &  16.4 \\ 
8084.477 &  -7.5 &  0.64 &  11.1 &  18.0 &  20.0 & 8514.336 & -10.0 &  1.14 &  12.3 &  14.8 &  16.1 & 9177.449 &  -9.8 &  0.88 &   3.3 &   7.1 &   6.7 \\ 
8100.455 &  -8.0 &  0.64 &  10.7 &  19.4 &  22.2 & 8520.295 & -11.7 &  1.26 &  10.7 &  10.2 &  11.5 & 9178.453 & -10.0 &  0.91 &   3.9 &   5.1 &   7.1 \\ 
8106.489 &  -8.2 &  0.68 &  13.8 &  15.0 &  14.4 & 8520.398 & -12.0 &  1.29 &   8.5 &   9.2 &  10.1 & 9181.475 &  -9.8 &  0.89 &   0.6 &   3.7 &   5.5 \\ 
8128.308 &  -7.7 &  0.67 &   2.5 &   9.2 &   9.3 & 8522.784 & -11.4 &  1.24 &   6.5 &   7.1 &   8.2 & 9182.465 &  -9.9 &  0.92 &   0.7 &   8.1 &   4.9 \\ 
8130.409 &  -8.9 &  0.70 &  -1.0 &  13.6 &  13.8 & 8524.446 & -12.2 &  1.31 &  12.9 &  13.2 &  13.2 & 9188.412 &  -9.6 &  0.78 &   8.6 &  12.2 &  13.2 \\ 
8132.300 &  -9.2 &  0.77 &  16.2 &  20.2 &  22.2 & 8528.416 & -12.1 &  1.32 &  18.7 &  16.5 &  17.3 & 9218.495 & -10.3 &  0.86 &  10.2 &  15.2 &  15.7 \\ 
8132.353 &  -9.2 &  0.75 &   8.5 &  15.4 &  17.6 & 8529.315 & -12.4 &  1.32 &  16.6 &  12.7 &  15.7 & 9220.482 & -10.3 &  0.88 &  12.8 &  17.2 &  17.2 \\ 
8137.370 &  -8.5 &  0.75 &  -6.6 &   7.6 &   8.8 & 8538.261 & -12.5 &  1.43 &  12.1 &   7.2 &  10.5 & 9223.302 & -11.2 &  0.88 &   5.2 &  13.1 &  13.8 \\ 
8138.542 &  -8.3 &  0.71 &  13.2 &  18.3 &  21.8 & 8540.271 & -11.9 &  1.37 &  16.3 &  14.7 &  16.5 & 9230.371 & -10.7 &  0.90 &   6.7 &  16.0 &  15.8 \\ 
8139.413 &  -9.0 &  0.73 &  12.4 &  19.2 &  20.8 & 8541.435 & -12.4 &  1.45 &  12.1 &  10.8 &  12.2 & 9230.620 & -10.1 &  0.88 &  13.3 &  24.2 &  25.0 \\ 
8141.447 &  -8.6 &  0.74 &  13.0 &  16.3 &  14.9 & 8542.741 & -11.5 &  1.34 &  11.9 &  12.6 &  13.2 & 9260.286 & -11.7 &  0.83 &  14.0 &  25.4 &  24.6 \\ 
8143.319 &  -9.4 &  0.76 &  11.9 &  18.7 &  18.8 & 8551.704 & -10.3 &  1.27 &  13.6 &  14.3 &  14.3 & 9269.661 & -10.5 &  0.68 &   9.5 &  23.1 &  22.6 \\ 
8143.470 &  -9.4 &  0.73 &   7.3 &  17.0 &  19.9 & 8559.745 & -10.2 &  1.19 &  16.2 &  17.9 &  17.3 & 9272.366 & -11.2 &  0.74 &  12.3 &  25.6 &  25.7 \\ 
8147.444 &  -9.5 &  0.75 &   3.1 &  17.1 &  18.9 & 8564.339 &  -9.0 &  1.12 &  31.3 &  30.1 &  30.8 & 9274.281 & -11.1 &  0.77 &  14.3 &  18.6 &  18.2 \\ 
8150.385 &  -9.8 &  0.82 &   2.6 &  17.0 &  19.9 & 8564.554 &  -9.7 &  1.15 &  20.6 &  21.2 &  20.7 & 9274.434 & -11.1 &  0.78 &  14.8 &  25.4 &  26.2 \\ 
8151.515 &  -9.0 &  0.74 &  14.4 &  24.5 &  24.6 & 8567.768 & -10.0 &  1.21 &  19.1 &  24.9 &  26.0 & 9276.365 & -11.3 &  0.78 &   9.4 &  20.0 &  19.4 \\ 
8155.309 &  -9.2 &  0.76 &  11.8 &  18.7 &  18.4 & 8568.326 & -11.0 &  1.28 &  20.8 &  25.6 &  25.9 & 9313.356 & -13.4 &  1.21 &   2.0 &  11.2 &  13.1 \\ 
8155.466 &  -8.8 &  0.74 &  21.7 &  26.1 &  25.5 & 8572.323 & -10.1 &  1.13 &  21.0 &  25.4 &  25.5 & 9315.641 & -15.0 &  1.29 &   6.8 &  14.1 &  16.9 \\ 
8157.378 &  -9.8 &  0.77 &  14.2 &  22.9 &  25.3 & 8578.675 &  -8.9 &  1.04 &  25.6 &  31.8 &  31.8 & 9515.556 & -15.8 &  2.02 &  25.9 &  28.8 &  29.7 \\ 
8158.534 &  -9.2 &  0.75 &  17.2 &  23.2 &  20.8 & 8581.677 &  -9.2 &  1.04 &  23.0 &  29.5 &  30.0 & 9531.722 & -14.9 &  1.83 &  27.1 &  28.5 &  31.3 \\ 
8166.316 &  -9.7 &  0.91 &  19.7 &  28.4 &  27.8 & 8583.690 &  -9.4 &  1.00 &  19.6 &  22.8 &  24.6 & 9533.736 & -15.4 &  1.89 &  22.3 &  24.4 &  27.3 \\ 
8167.310 & -10.1 &  0.93 &  21.2 &  33.5 &  32.6 & 8584.692 &  -9.3 &  0.97 &  26.2 &  31.1 &  31.0 & 9534.749 & -14.7 &  1.82 &  24.9 &  26.6 &  27.7 \\ 
8167.453 &  -9.6 &  0.84 &  15.1 &  24.3 &  23.1 & 8586.367 &  -9.5 &  1.05 &  24.3 &  27.1 &  26.2 & 9539.480 & -14.8 &  1.82 &  26.2 &  26.5 &  29.9 \\ 
8171.387 & -10.3 &  1.02 &  16.8 &  26.2 &  26.3 & 8586.678 &  -9.3 &  1.01 &  21.7 &  28.2 &  29.2 & 9542.451 & -14.4 &  1.70 &  23.7 &  23.2 &  25.9 \\ 
8174.497 &  -9.8 &  1.07 &  16.0 &  28.0 &  27.9 & 8595.687 &  -8.1 &  0.89 &  26.1 &  32.1 &  34.1 & 9551.493 & -14.2 &  1.59 &  26.1 &  25.4 &  27.0 \\ 
8195.579 & -10.6 &  1.43 &  16.9 &  29.5 &  29.9 & 8755.682 & -15.5 &  2.01 &  17.9 &  21.7 &  24.0 & 9551.675 & -13.5 &  1.49 &  22.2 &  22.1 &  26.2 \\ 
8198.320 & -10.2 &  1.42 &  22.0 &  31.1 &  32.0 & 8764.668 & -16.8 &  2.16 &  12.8 &  16.6 &  16.0 & 9562.434 & -15.0 &  1.55 &  16.0 &  14.4 &  15.9 \\ 
8198.359 & -10.9 &  1.44 &  13.9 &  26.5 &  27.3 & 8783.624 & -15.2 &  2.12 &   8.6 &   8.0 &   8.5 & 9563.431 & -14.3 &  1.49 &  16.6 &  12.9 &  17.1 \\ 
8199.303 & -10.7 &  1.44 &  12.3 &  24.8 &  26.5 & 8808.537 & -12.4 &  1.76 &  14.0 &  15.6 &  16.3 & 9568.655 & -13.4 &  1.49 &  16.7 &  14.5 &  16.7 \\ 
8199.394 & -10.0 &  1.36 &  18.4 &  26.6 &  27.4 & 8820.698 & -13.6 &  1.83 &  18.9 &  20.9 &  20.8 & 9582.672 & -13.7 &  1.51 &  13.2 &  11.3 &  12.5 \\ 
8213.646 & -10.7 &  1.39 &  24.4 &  26.9 &  28.7 & 8836.372 & -12.4 &  1.56 &  25.2 &  31.9 &  33.3 & 9593.309 & -12.1 &  1.35 &  16.2 &  14.1 &  16.2 \\ 
8222.368 & -10.7 &  1.30 &  22.6 &  29.9 &  32.3 & 8838.481 & -12.0 &  1.57 &  22.1 &  24.4 &  25.3 & 9596.466 & -12.6 &  1.38 &  20.1 &  17.0 &  17.2 \\ 
8223.323 & -10.9 &  1.29 &  22.0 &  26.5 &  27.3 & 8845.517 & -12.1 &  1.48 &  20.5 &  25.8 &  26.6 & 9600.548 & -13.1 &  1.31 &  17.8 &  15.3 &  16.1 \\ 
8232.338 & -10.6 &  1.27 &  15.9 &  24.0 &  25.9 & 8848.437 & -11.6 &  1.39 &  17.0 &  20.9 &  22.4 & 9604.381 & -13.1 &  1.31 &  15.6 &  11.1 &  12.2 \\ 
8234.605 &  -9.9 &  1.23 &  21.0 &  21.9 &  23.4 & 8855.503 & -11.3 &  1.34 &  19.6 &  25.8 &  27.1 & 9604.413 & -13.7 &  1.31 &  17.7 &  14.7 &  14.1 \\ 
8358.681 &  -7.4 &  0.80 &  24.5 &  29.3 &  30.8 & 8864.304 & -10.0 &  1.15 &  25.2 &  28.4 &  28.6 & 9613.605 & -11.3 &  1.08 &  16.1 &  16.6 &  15.9 \\ 
8362.668 &  -7.2 &  0.78 &  19.9 &  27.1 &  28.3 & 8867.472 & -11.0 &  1.20 &  19.9 &  23.9 &  23.6 & 9618.402 & -11.9 &  1.09 &  19.4 &  20.5 &  20.9 \\ 
8363.581 &  -7.5 &  0.78 &  16.9 &  18.7 &  18.0 & 8868.282 & -10.6 &  1.21 &  17.3 &  22.2 &  23.0 & 9628.595 & -11.7 &  1.04 &  24.4 &  26.3 &  28.2 \\ 
8365.641 &  -6.8 &  0.72 &  19.3 &  18.1 &  16.6 & 8872.422 & -10.7 &  1.11 &  24.2 &  31.2 &  31.5 & 9636.681 & -11.5 &  1.01 &  21.7 &  26.8 &  28.3 \\ 
8369.601 &  -6.8 &  0.71 &  11.1 &  20.0 &  21.4 & 8884.387 &  -9.9 &  0.99 &  16.4 &  16.7 &  16.3 & 9637.292 & -11.3 &  1.02 &  10.9 &  16.7 &  17.1 \\ 
8377.608 &  -7.3 &  0.71 &  11.6 &  19.8 &  19.0 & 8886.284 & -10.5 &  1.02 &  15.3 &  19.7 &  22.3 & 9637.402 & -12.2 &  1.03 &  16.6 &  22.6 &  24.8 \\ 
8388.695 &  -7.3 &  0.66 &   5.9 &   8.5 &  12.8 & 8886.362 & -11.4 &  1.05 &  18.7 &  20.1 &  21.5 & 9641.307 & -11.6 &  1.01 &  19.6 &  25.8 &  24.9 \\ 
\hline
  \end{tabular}

  {\scriptsize $EWs$ were evaluated  between --600\,km\,s$^{-1}$ and 600\,km\,s$^{-1}$; `depth' corresponds to the normalized amplitude of the lowest point in the central absorption ($<1$ if below continuum, $>1$ if above it); $RVs$ were evaluated by three methods (see text): the first order moment (column M1), the mirror method (column mirror), and the double-Gaussian method (column 2G).}
\end{table*}

\bsp	
\label{lastpage}
\end{document}